\documentclass[8.5pt,twoside,twocolumn]{article}
\usepackage[super,sort&compress,comma]{natbib} 
\usepackage{mhchem}
\usepackage{times,mathptmx}
\usepackage{sectsty}
\usepackage{balance} 
\usepackage{amsmath,fixltx2e} 
\usepackage{lastpage,color}
\usepackage{graphicx}
\usepackage[format=plain,justification=raggedright,singlelinecheck=false,font=small,labelfont=bf,labelsep=space]{caption}
\usepackage{fancyhdr}
\usepackage{epstopdf}
\fancypagestyle{plain}{
\fancyhead[L]{\includegraphics[height=8pt]{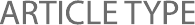}}
\fancyhead[C]{\hspace{-1cm}\includegraphics[height=20pt]{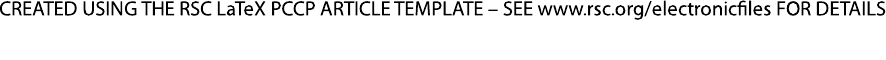}}
\fancyhead[R]{\includegraphics[height=10pt]{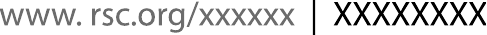}\vspace{-0.2cm}}
}

\makeatletter 
\def\subsubsection{\@startsection{subsubsection}{3}{10pt}{-1.25ex plus -1ex minus -.1ex}{0ex plus 0ex}{\normalsize\bf}} 
\def\paragraph{\@startsection{paragraph}{4}{10pt}{-1net.25ex plus -1ex minus -.1ex}{0ex plus 0ex}{\normalsize\textit}} 
\renewcommand\@biblabel[1]{#1}            
\renewcommand\@makefntext[1]%
{\noindent\makebox[0pt][r]{\@thefnmark\,}#1}
\makeatother 

\sectionfont{\large}
\subsectionfont{\normalsize} 

\begin{document}
\twocolumn[
  \begin{@twocolumnfalse}
\noindent\LARGE{\textbf{An electrical conductivity relaxation study of oxygen transport in  samarium doped ceria}}
\vspace{0.6cm}

\noindent\large{\textbf{Chirranjeevi Balaji Gopal\textit{$^{a}$} and
Sossina M. Haile$^{\ast}$\textit{$^{ab}$}}}\vspace{0.5cm}

\noindent\textit{\small{\textbf{Received Xth XXXXXXXXXX 20XX, Accepted Xth XXXXXXXXX 20XX\newline
First published on the web Xth XXXXXXXXXX 200X}}}

\noindent \textbf{\small{DOI: 10.1039/b000000x}}
\vspace{0.6cm}

\noindent \normalsize{The efficacy of the electrical conductivity relaxation (ECR) technique for investigating the oxygen transport properties of mixed conducting oxides has been evaluated. Fifteen mol\% samarium doped ceria (SDC15), for which approximate values of the two principal transport properties, bulk oxygen diffusivity, $D_{Chem}$, and surface reaction rate constant, $k_S$, can be found in the literature, was chosen as the benchmark material against which to validate the methodology. Measurements were carried out at temperatures between 750 $^\circ$C and 850 $^\circ$C and over a wide range of oxygen partial pressures. An unexpectedly high p-type electronic transference number enabled ECR measurements under oxidizing conditions. A systematic data analysis procedure was developed to permit reliable extraction of the kinetic parameters even in the general case of simultaneous bulk and surface limitation. The $D_{Chem}$ from this study showed excellent qualitative and quantitative agreement with expected values, falling in the range from $\sim 2\times10^{-5}$ to $2\times10^{-4}$ cm$^2$/s. The surface reaction constant under H$_2$/H$_2$O mixtures also showed good agreement with literature results. Remarkably, this value increased by a factor of 40 under mixtures of CO/CO$_2$ or O$_2$/Ar. This observation suggests kinetic advantages for production of CO rather than H$_2$ in a two-step solar-driven thermochemical process based on samarium doped ceria.}
\vspace{0.5cm}
 \end{@twocolumnfalse}
]

\footnotetext{\textit{$^{a}$ Materials Science, California Institute of Technology, USA. E-mail: smhaile@caltech.edu}}
\footnotetext{\textit{$^{b}$ Chemical Engineering, California Institute of Technology, USA.}}
\section{Introduction}
The remarkable capacity of ceria to display significant oxygen nonstoichiometry ($\delta$) at high temperatures or low oxygen activity without changing its crystal structure is essential to many of its applications in solid state electrochemistry. Beyond its widespread use as a solid-oxide fuel-cell electrolyte when doped with trivalent elements such as samarium or gadolinium, nonstoichiometric ceria (CeO$_{2-\delta}$) has recently emerged as a candidate reaction medium to facilitate two-step solar thermochemical splitting of water and/or carbon dioxide to generate hydrogen or other fuels\cite{Abanades2006,LeGal2011,Chueh2010,Chueh2010b,Chueh2009}. The first of the two steps is a high temperature endothermic reaction involving bulk release of oxygen. The second step, typically performed at a lower temperature, is the oxidation of the reduced ceria by the reactant gases (H$_2$O and/or CO$_2$) that returns the oxide to a low value of oxygen nonstoichiometry. \\

Whereas thermodynamics governs the theoretically achievable fuel productivity from this pair of reactions, that is, the fuel produced per cycle, the rate at which fuel is produced, the other critical metric, is a function of kinetics. Two serial steps are involved: diffusion of neutral oxygen species within the bulk of the oxide, quantified in terms of the chemical diffusion coefficient $D_{Chem}$, and reaction at the surface of the oxide, quantified in terms of the surface reaction rate constant $k_S$. In principle, $D_{Chem}$ and $k_S$ are embodied in the time evolution of oxygen release or fuel production in a thermochemical experiment. In practice, however, the large driving forces (\textit{i.e.} large changes in \textit{T} and \textit{p}O$_2$), the random porous microstructure of the materials commonly employed, and the poorly controlled gas flow dynamics of the typical thermochemical reactors 
preclude access to these terms and impede meaningful comparisons of the kinetic responses of candidate materials. In contrast to fuel production studies, experiments aimed at directly and quantitatively revealing the kinetic properties must use small perturbations from equilibrium to avoid complex, non-linear effects, must employ well-defined sample geometries, and must present well-controlled gas flow dynamics.\\

A variety of techniques have been employed in combination with experimental configurations that meet the above requirements for measuring $D_{Chem}$ and $k_S$. These include secondary ion mass spectrometry (SIMS) to analyze isotope depth profiles\cite{Lane2000a}, gravimetry relaxation\cite{Katsuki2002,Yashiro2002}, electrochemical impedance spectroscopy\cite{Lai2005} and electrical conductivity relaxation\cite{Yasuda1994,Jacobson2000,Song1999}. The objective of the present work is to demonstrate the versatility of this last method, electrical conductivity relaxation (ECR), to study the effect of temperature and gas atmosphere on $D_{Chem}$ and $k_S$. \\

In a relaxation experiment, one analyzes transient behavior in the re-equilibration process following a step change in the \textit{p}O$_2$ of the surrounding gas. The relaxation profile, typically that of sample mass or electrical conductivity, is described by a solution to Fick's second law that takes into account the appropriate boundary conditions. A fit to the data yields values for the desired material parameters. The conductivity relaxation method is particularly attractive because of the ease with which electrical conductivity can be measured and with which reactors with small volumes, as required for rapid exchange of gases, can be constructed. The long history of the ECR method, having been practiced as early as 1934 by D\"{u}nwald and Wagner\cite{Wagner1934} renders the technique, in some sense, a `classic' tool. Furthermore, in some quarters, the level of sophistication in its application has yielded highly compelling results\cite{Song1999}. In many other instances, however, the experimental and numerical requirements for the success of the method are not fully appreciated.  Indeed, it has been recently suggested that a simultaneous determination of $D_{Chem}$ and $k_{S}$ is inherently unreliable\cite{Cox-galhotra2010}.\\

In the present study we have performed ECR measurements on bulk samples of Sm$_{0.15}$Ce$_{0.85}$O$_{1.925-\delta}$ (samaria doped ceria, SDC15) to extract both $D_{Chem}$ and $k_{S}$ with the dual objectives of demonstrating the conditions under which both parameters can be reliably determined and providing new insights into this technologically important oxide. SDC15 is an ideal material against which to validate the experimental and analytical methodologies because the bulk transport properties are well-known and, though to a lesser degree of certainty, the surface properties are also known\cite{Lai2005}. In addition, despite significant interest in SDC, surprisingly, comprehensive studies of its surface reactivity remain to be reported. Reports to date have either encompassed a limited range of oxygen partial pressures\cite{Katsuki2002} or have focused on phenomena such as the influence of bulk grain boundaries\cite{Wang2012}, thin-film thickness effects\cite{Karthikeyan2008}, or the role of metal/oxide interfaces\cite{Wang2013}, each under a narrow range of conditions. \\

This paper is organized as follows. Section~\ref{sec:theory} will briefly review the relevant theory for relaxation experiments and present a very brief analysis of anticipated results based on literature measurements of D$_{Chem}$ and k$_{S}$ in SDC15. In Section~\ref{sec:expt}, the experimental details will be presented, followed by our data analysis procedure and its test results. We will then discuss our results with SDC15 in Section~\ref{sec:results} before concluding with Section~\ref{sec:conclusions}. 
 
\section{Theory}\label{sec:theory}
\subsection{Electrical conductivity relaxation}
A detailed formulation of the diffusion model underlying the ECR method and its numerical analysis can be found in the literature\cite{denOtter2000,denOtter2001,Boukamp2004}. For completeness, we provide a brief theoretical background and highlight pertinent equations along with the key assumptions.\\

The sample geometry employed here is that of an infinite sheet of thickness `$2a$' along the direction, $x$, of oxygen transport. In response to the step change in gas phase oxygen partial pressure, the oxygen concentration varies with $x$ and with time, $t$. The conductivity, taken to be directly proportional to the oxygen concentration, is measured along a direction normal to that of oxygen transport. Solving Fick's second law of diffusion in 1D under the assumption that the surface reaction is first order in concentration with rate constant $k_S$, \textit{i.e.}
\begin{equation}\label{eq:BC}
J(\pm a) = \mp k_S(c_V(\pm a,t) - c_V(\pm a,\infty))\text{,}
\end{equation}
results in the following concentration profile\cite{Crank1975}:
\small
\begin{equation}
\frac{c_V(x,t) - c_V(0)}{c_V(\infty) - c_V(0)} = 1 - \sum_{m=1}^{\infty} \frac{2\tilde{L} \cos(\alpha_mx/a)}{(\alpha_m^2 + \tilde{L}^2 +\tilde{L}) \cos(\alpha_m)}\exp\left({-\frac{\alpha_m^2D_{Chem}t}{a^2}}\right) 
\end{equation}
\normalsize
where $c_V(\pm a,t)$ and $c_V(\pm a,\infty)$ are, respectively, the instantaneous and final volumetric concentrations of vacancies at the sample surface, and $\{\alpha_m\}$ is the set of positive roots of 
\small 
\begin{equation}\label{eq:constraint}
\alpha_m tan(\alpha_m) =\tilde{L} = \frac{ak_S}{D_{Chem}}, 
\end{equation}
\normalsize
where $\tilde{L}$ is a dimensionless length that reflects the relative roles of surface reaction and bulk diffusion in the overall relaxation rate.
Under the assumption of a total conductivity that varies linearly with concentration (valid when step changes in oxygen partial pressure are sufficiently small) the spatially averaged, normalized conductivity obtained from the measurement is 
\small 
\begin{equation}\label{eq:diffusion-surface}
\dfrac{\sigma(t)-\sigma(0)}{\sigma(\infty) -\sigma(0)} = 1 -\sum_{m=1}^{\infty} \dfrac{2\tilde{L}^2}{\alpha_m^2(\alpha_m^2+\tilde{L}^2 + \tilde{L})} \exp\left({-\frac{\alpha_m^2D_{Chem}t}{a^2}}\right) 
\end{equation}
\normalsize
where, $\sigma(0)$ and $\sigma(\infty)$ are, respectively, the initial and final equilibrated conductivities of the sample. \\

The form of the dimensionless conductivity is simplified under conditions in which only one process dominates. When the surface reaction step is much slower than bulk diffusion, \textit{i.e.}, $k_S \ll D_{Chem}/a$ and $\tilde{L}\ll1$, Equation~\ref{eq:constraint} becomes
\begin{equation}
\tilde{L} = \alpha_1\tan(\alpha_1) \approx \alpha_1^2 
\end{equation}
with
\begin{equation}
\alpha_m \approx m\pi \:\textrm{(m $\ge$ 2)}
\end{equation} 
This causes all but the first exponential in Equation~\ref{eq:diffusion-surface} to reduce to zero, such that
\begin{equation}\label{eq:surface}
\vspace*{.2cm}
\dfrac{\sigma(t)-\sigma(0)}{\sigma(\infty) -\sigma(0)} = 1- \exp\left(\frac{k_St}{a}\right) .
\end{equation}
At the other extreme of bulk diffusion limited transport, \textit{i.e.}, $k_S \gg D_{Chem}/a$ and thus $\tilde{L}\gg1$, the roots to Equation~\ref{eq:constraint} are 
\begin{equation}
\alpha_m = \frac{(2m-1)\pi}{2},
\end{equation}
and Equation~\ref{eq:diffusion-surface} becomes
\small
\begin{equation}\label{eq:diffusion}
\dfrac{\sigma(t)-\sigma(0)}{\sigma(\infty) -\sigma(0)} = 1 - \frac{64}{\pi^2} \sum_{m=1}^{\infty}\frac{1}{(2m-1)^2}\exp\left(-\frac{(2m-1)^2 \pi^2D_{Chem}t}{4a^2}\right) .
\end{equation}
\normalsize
The challenges associated with attempting to fit Equation~\ref{eq:diffusion-surface} to experimental data so as to determine the kinetic parameters have been addressed by many authors\cite{Song1999,Boukamp2004,Ciucci2013}. Because $\tilde{L}$ is not known \textit{a priori}, any one of Equations~\ref{eq:diffusion-surface}, \ref{eq:surface} or \ref{eq:diffusion} could potentially describe the relaxation profile. This necessitates a data analysis procedure that can reliably extract the parameters without under or over fitting. \\

Experimentally, success of the ECR method requires that several conditions be met. First, there must be no open porosity (which would allow gas phase access to the interior and greatly speed the relaxation process) and minimal closed porosity (which would slightly retard the process by limiting bulk diffusion). Second, the reactor flush time ($t_0$) must be much smaller than the material response time ($\tau$), where $\tau$ is $\approx a/k_S$ in the surface reaction limited regime and $\approx a^2/4D_{Chem}$ in the diffusion limited regime. Third, the grain sizes must be large (on the order of microns) so as to minimize grain boundary contributions to the measured electrical resistance and also to eliminate possibilities of a grain-boundary mediated relaxation process. The latter, while certainly of significant scientific interest, would render Equation~\ref{eq:diffusion-surface} inapplicable. Finally, the step changes must be made small to validate the assumption of first order surface reaction kinetics and constant $D_{Chem}$ and $k_S$ between the initial and final \textit{p}O$_2$ values. This also guarantees that the magnitude of the thermodynamic driving force is the same regardless of the direction of \textit{p}O$_2$ change and justifies the assumption that conductivity varies linearly with oxygen content. Exactly how small the step change must be depends on the details of the system under investigation and has been discussed at length by Jacobson and co-workers\cite{Wang2001}. In the present study, $\Delta \ln(p\text{O}_2$) was restricted to a value of $<$ 0.5, and the sufficiency of this choice was evaluated by comparing results for the forward and backward step changes.\\

A fourth requirement, specific to the geometry described here, is that the width $2a$ of the sample along the direction of interest must be much smaller than those of the other two directions to justify the 1-dimensional solution. In principle, solutions that treat the 2- and 3-dimensional cases are available\cite{Crank1975}, and even in these cases only two free parameters are obtained from the fit. In practice, however, ensuring that a global minimum is obtained is much more challenging. In this work some measurements were carried out using samples that, in fact, necessitated analysis according to the 2-D solution, as discussed at the end of Section~\ref{subsec:num_details}. 

\subsection{Defect chemistry and conductivity of SDC15}
The defect chemical origins of \textit{p}O$_2$-dependent conductivity in rare-earth doped ceria are well established\cite{Mogensen2000} and are briefly reviewed here for completeness. When the dopant concentration is high relative to the intrinsic defect concentrations, global electroneutrality reduces to 
\begin{equation}\label{eq:electroneutrality}
[\text{Sm}_{\text{Ce}}^\prime]=2[\text{V}_\text{O}^{\bullet\bullet}]
\end{equation}
where Kr\"{o}ger-Vink notation\cite{Kroger1956} has been employed, and $[\text{Sm}_\text{Ce}^\prime]$ and $[\text{V}_\text{O}^{\bullet\bullet}]$ are, respectively, the fractional dopant and oxygen vacancy concentrations. Despite an approximately fixed vacancy concentration, the mobile electron concentration in ceria varies with oxygen chemical potential, \textit{i.e.}, \textit{p}O$_2$, as a result of the reduction reaction.
\begin{equation}\label{eq:defect_reaction}
\text{O}_\text{O}^\text{x} \longleftrightarrow \frac{1}{2}\text{O}_2(g) + \text{V}_\text{O}^{\bullet\bullet} + 2e^\prime
\end{equation}
with an equilibrium constant $K_R(T)$,
\begin{equation}\label{eq:equilibrium_constant}
K_R(T)\approx[\text{V}_\text{O}^{\bullet\bullet}]n^2p\text{O}_2^{1/2} \approx 1/2[\text{Sm}_\text{Ce}^{\prime}]n^2p\text{O}_2^{1/2}
\end{equation}
where $n$ represents the fractional concentration of mobile electrons and describes equally well the Ce$^{3+}$ concentration. In addition, thermal excitation generates electrons (equivalently, Ce$^{3+}$) and holes (equivalently, O$^-$) from Ce$^{4+}$ and O$^{2-}$ species \cite{Yokokawa2006}, the concentrations of which can be expected to obey the relationship 
\begin{equation}\label{eq:e-h-pair}
np = K_{np}(T)
\end{equation}
These expressions, in combination with the much higher mobility of electronic defects than ionic defects, give rise in principle to a \textit{p}O$_2$-dependent conductivity of the form
\begin{equation}\label{eq:conductivity_model_expected}
\sigma_{tot} = \sigma_n + \sigma_{ion} + \sigma_p = \sigma_n^0p\text{O}_2^{-1/4} +\sigma_{ion} + \sigma_p^0p\text{O}_2^{1/4}
\end{equation}
where $\sigma_n^0$ and $\sigma_p^0$ are constants that depend on the dopant concentration, the reduction equilibrium constant, the electronic defect equilibrium constant and the respective electronic defect mobilities. This relationship implies a double-logarithmic plot of $\sigma_{tot}$ vs \textit{p}O$_2$ will display a flat region, reflecting the electrolytic domain, which is flanked by regions at low and high \textit{p}O$_2$ with slopes of -1/4 and 1/4, respectively, corresponding to the n-type and p-type regimes. \\

Numerous experimental measurements of total conductivity have revealed the existence of the electrolytic and ideal n-type domains in doped ceria\cite{Wang2000,Lai2005}. In contrast, the p-type conductivity, which is generally lower than the n-type conductivity over accessible $p\text{O}_2$ ranges due to the much lower concentration of holes, has been reported only on the basis of partial conductivity measurements\cite{Wiemhofer1998,Xiong2004,Xiong2008}. The question naturally arises, then, whether the variation in conductivity under oxidizing conditions ($p\text{O}_2>10^{-5}$ atm) is sufficient to permit a meaningful ECR measurement. Based on the p-type conductivity measured by Xiong \textit{et al}.\cite{Xiong2004} for SDC20 at 800 $^\circ$C and the ionic conductivity of SDC20 reported by Yahiro \textit{et al}.\cite{Yahiro1988} at the same temperature, one can estimate that the relative change in conductivity on changing the gas atmosphere from 1 to 0.1 atm \textit{p}O$_2$ will be on the order of 0.3\% (with an absolute conductivity on the order of 0.032 S/cm). Achieving sensitivity at this level, though requiring care, is not prohibitive. Accordingly, and because the surface reaction properties of doped ceria under oxidizing conditions are as important for thermochemical cycling as are the properties under reducing conditions, measurements were made under a wide \textit{p}O$_2$ range, including the oxidizing regime.

\subsection{Mass transport : chemical diffusivity and surface reactivity}
The chemical or ambipolar diffusion coefficient in a mixed conducting oxide describes the concerted flux of oxide ion and electronic defects under an oxygen chemical potential gradient\cite{Maier2005}. In the dilute limit, $D_{Chem}$ can be expressed as a function of the ionic conductivity, $\sigma_{ion}$, the electronic conductivity, $\sigma_e$, and the corresponding volumetric defect concentrations, $c_{ion}$ and $c_{e}$, as follows\cite{Maier1993}
\begin{equation}\label{eq:D-ideal}
D_{Chem} =  \frac{RT}{4F^2}\frac{\sigma_{ion}\sigma_e}{\sigma_{ion}+\sigma_e}\left[\frac{1}{c_{ion}} + \frac{4}{c_e}\right],
\end{equation}
where F and R are Faraday's constant and the universal gas constant, respectively. In a material such as SDC15, oxygen vacancies are unquestionably the relevant ionic defects ($c_{ion} = c_V$), whereas under conditions of negligible hole conductivity, the electronic defects of relevance are the mobile electrons ($\sigma_{e} = \sigma_n$ and $c_{e} = c_n$). Thus, with knowledge of the conductivities and concentrations of these two types of carriers, the ambipolar diffusion coefficient can be computed.\\

As already discussed in the context of the defect chemistry, conductivity is often directly measured, and for SDC15 both $\sigma_{ion}$ and $\sigma_n$ are readily available in the literature as functions of temperature and, in the latter case, of \textit{p}O$_2$ as well. The remaining unknowns, the defect concentrations, are obtained by noting that, within the electroneutrality regime defined by Equation~\ref{eq:electroneutrality}, the vacancy concentration is, by definition, fixed by the dopant concentration. The electron concentration is implied by Equation~\ref{eq:equilibrium_constant}, which on rearrangement and combination with Equation~\ref{eq:electroneutrality}, becomes\cite{Mogensen2000}
\begin{equation}
n = \left(\frac{2K_R(T)}{[\text{Sm}_\text{Ce}^\prime]}\right)^{1/2}p\text{O}_2^{-1/4}
\end{equation}

The equilibrium reduction constant for SDC15 has been reported in the literature, and the individual thermodynamic terms, the entropy, $\Delta S_O$, and enthalpy, $\Delta H_O$, of reduction, which give $K_R$ according to
\begin{equation}
K_R(T) = \exp\left( \frac{\Delta S_O}{k_B} \right)\exp\left(\frac{-\Delta H_O}{k_BT}\right)
\end{equation}
are available\cite{Lai2005}.
Thus, using literature values for $\sigma_{ion}$, $\sigma_n$, $\Delta H_O$, $\Delta S_O$, and the molar volume to convert from fractional to volumetric defect concentrations, it is possible to compute $D_{Chem}$, against which experimental results for $D_{Chem}$ can be compared. Indeed, directly measured values of $D_{Chem}$ have generally shown good agreement with those computed according to Equation~\ref{eq:D-ideal}\cite{Yashiro2002}. \\

Turning to the transport across the gas-solid interface, the surface reaction rate constant in doped ceria has also been evaluated in the literature, not only using relaxation methods\cite{Yashiro2002,Katsuki2002}, but also using A.C. impedance spectroscopy (ACIS)\cite{Lai2005} and oxygen isotope exchange measurements\cite{Yashiro2002,Kilner2000}. In an impedance measurement, one typically obtains an area-normalized electrochemical (or electrode) resistance term, $\rho_{electrode}$, often referred to simply as the ‘area-specific-resistance’ or ASR. For a surface active oxide (in contrast to one that is electrochemically active only at the triple phase boundaries formed between the oxide, metal and gas phase) this resistance implies a surface reaction constant defined according to \cite{Lai2005}
\begin{equation}\label{eq:rhoelectrode}
k_S= \frac{k_BT}{(ze)^2\rho_{electrode}c_{V}},
\end{equation}
where $e$ is the elementary charge, $z$ is the valence of the species (2 for oxygen vacancies) and $k_B$ is Boltzmann's constant. Formally, the vacancy concentration in Equation~\ref{eq:rhoelectrode} is that at the surface, but in the absence of detailed knowledge of the surface characteristics, $c_{V}$ can be reasonably approximated by the bulk value. Furthermore, because of the equivalence between charge and mass transport across the interface, this electrochemically determined reaction constant is identical to the surface reaction constant obtained from ECR measurements\cite{Maier2005}. \\

In contrast to the direct equivalence between surface reaction constants obtained from ECR and ACIS methods, the surface exchange constant obtained from isotope exchange measurements, $k_S^{ex}$, is related to the former terms by a proportionality constant that depends on the material thermodynamic behavior. Specifically, it can be shown that \cite{Kim2000,Yashiro2002}
\begin{equation}
k_S = k_{S}^{ex}\frac{\partial \ln a_O}{\partial \ln [\text{O}_\text{O}^\times]}
\end{equation}
where $a_O$ and [O$_\text{O}^\times$], are, respectively, the activity and concentration of oxygen atoms in the bulk of ceria. In the dilute limit, $a_O=[\text{O}_\text{O}^\times]$ and the two rate constants become equal. In light of the many methods available for determining the surface reaction constant, it is not surprising then that there are several experimental reports\cite{Lane2000a,Katsuki2002,Lai2005} against which the values measured here can be compared.\\

In addition to method validation, approximate values of $D_{Chem}$ and $k_S$ from the literature permit a preliminary identification of the rate-limiting step for a given sample thickness. Specifically, the critical thickness, $L_c = D_{Chem}/k_S$, delineates the surface and bulk limited regimes in that samples with $a < L_c$ are largely surface reaction limited and conversely those with $a > L_c$ are largely bulk diffusion limited\cite{Crank1975}. For 10-20 mole\% rare-earth doped ceria, reported $D_{Chem}$ values range from $2\times10^{-5}$ to $1\times10^{-4}$ cm$^2$/s at temperatures from 750 to 850 $^\circ$C and oxygen partial pressures from $10^{-24}$ atm to $10^{-3}$ atm. Typical values of $k_S$ from ECR and impedance measurements under similar conditions are on the order of $5\times10^{-6}$ cm/s to $1\times10^{-5}$ cm/\cite{Katsuki2002,Lai2005}s. Taking $D_{Chem} \approx 1\times10^{-5}$ cm$^2$/s and $k_S \approx 1\times10^{-5}$ cm/s yields $L_c\approx1$ cm. Thus, a typical sample of thickness 0.8 mm as used in these experiments can be expected to be well within the surface-reaction limited regime.

\section{Experimental and Analytical Procedure}\label{sec:expt}
\subsection{Experimental methods}\label{subsec:expt_details}
\begin{figure}[!h]
\includegraphics[scale=0.35]{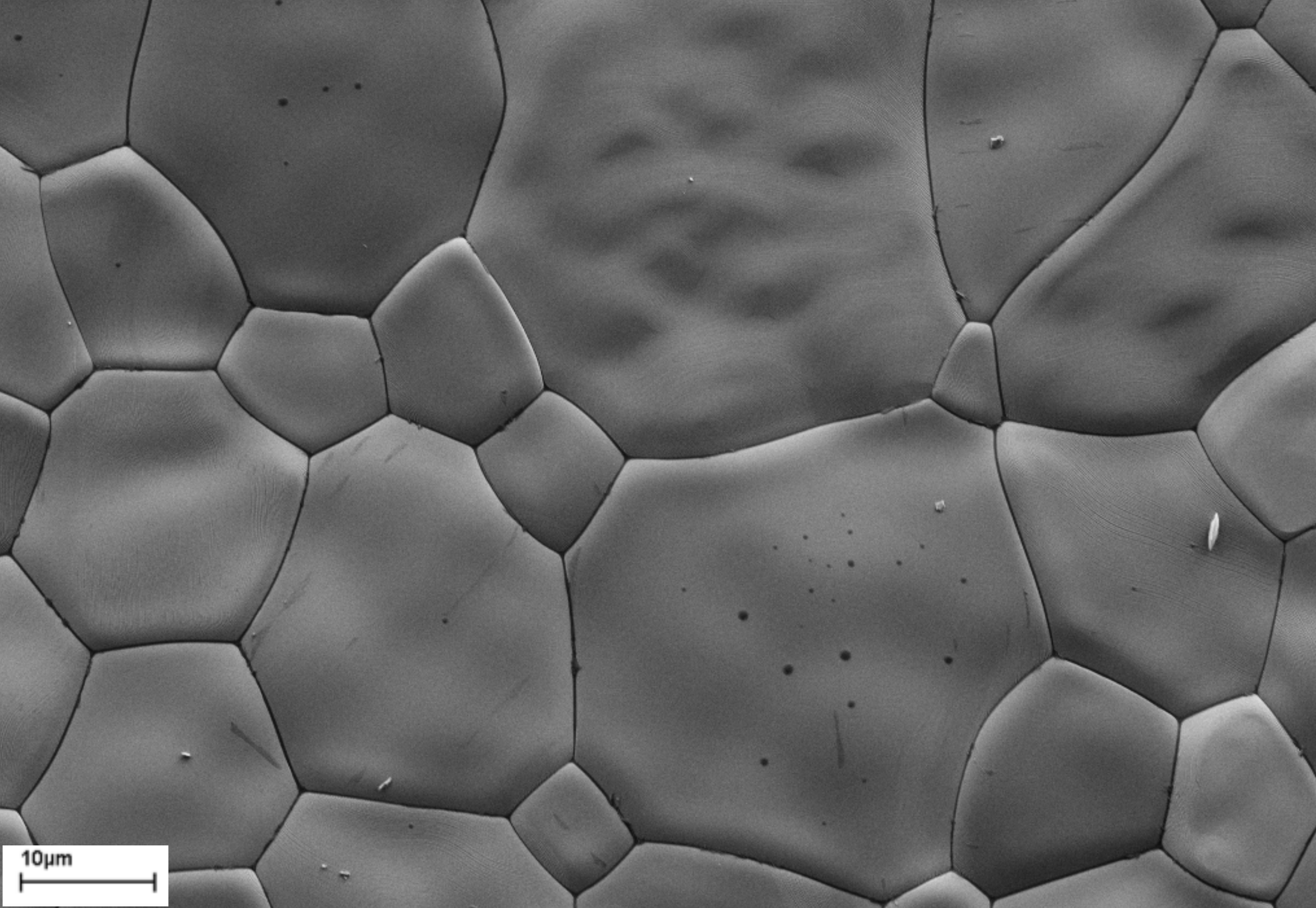}
\caption{Scanning electron micrograph of a sintered SDC15 pellet (unpolished) showing average grain size of 3 microns and minimal porosity.}
\label{fig:sample_microstructure}
\end{figure}

 \begin{figure}[!h]
 \centering
\includegraphics[scale=0.35]{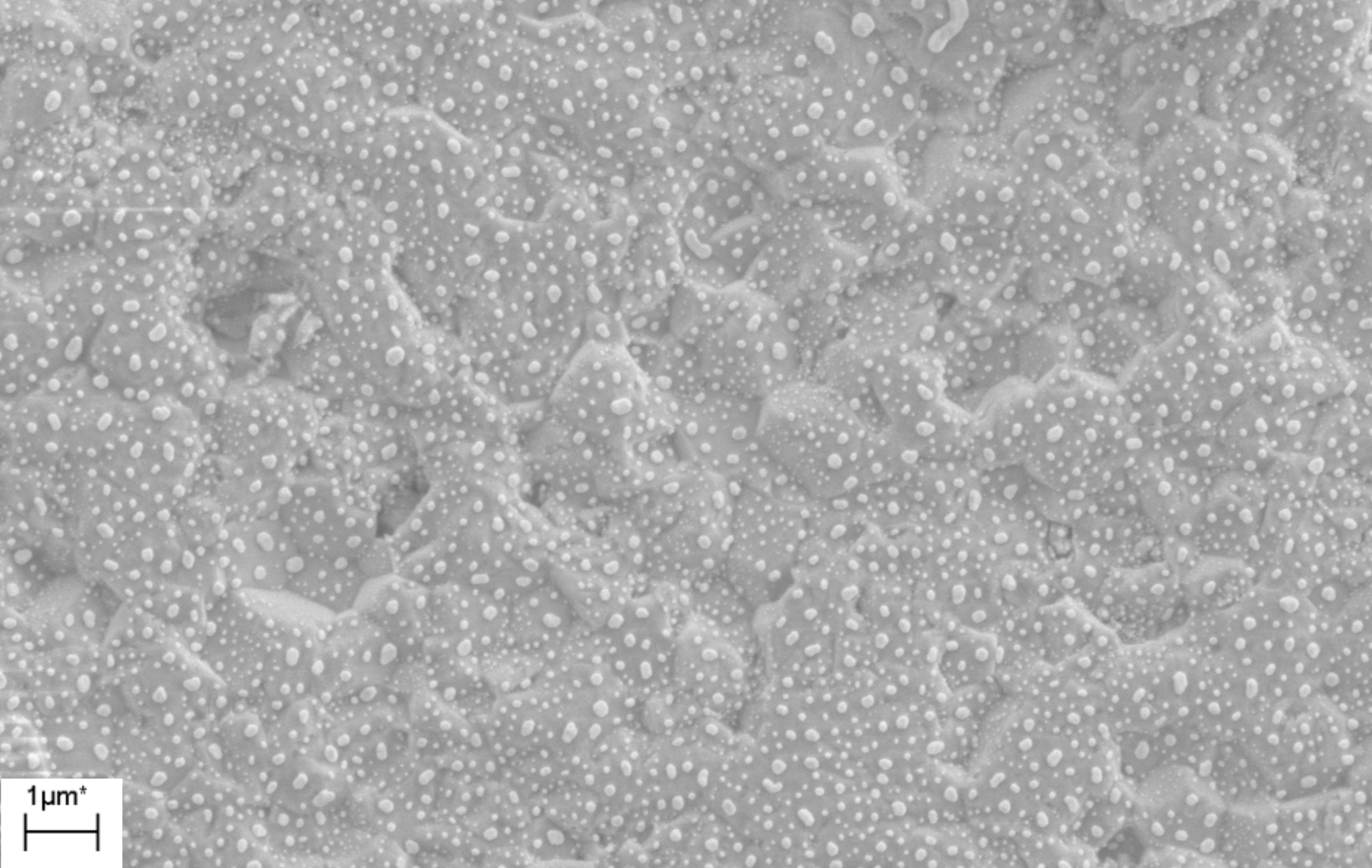}
\caption{Scanning electron micrograph showing isolated but well dispersed Pt catalyst particles sputtered on an SDC15 sample and  annealed at 950 $^\circ$C for an hour. The average particle size was close to 100 nm, with an interparticle spacing of 400 nm.}
\label{fig:Pt-particles}
\end{figure}

\begin{figure*}[!h]
\hspace{-0.4cm}\includegraphics[scale=0.8]{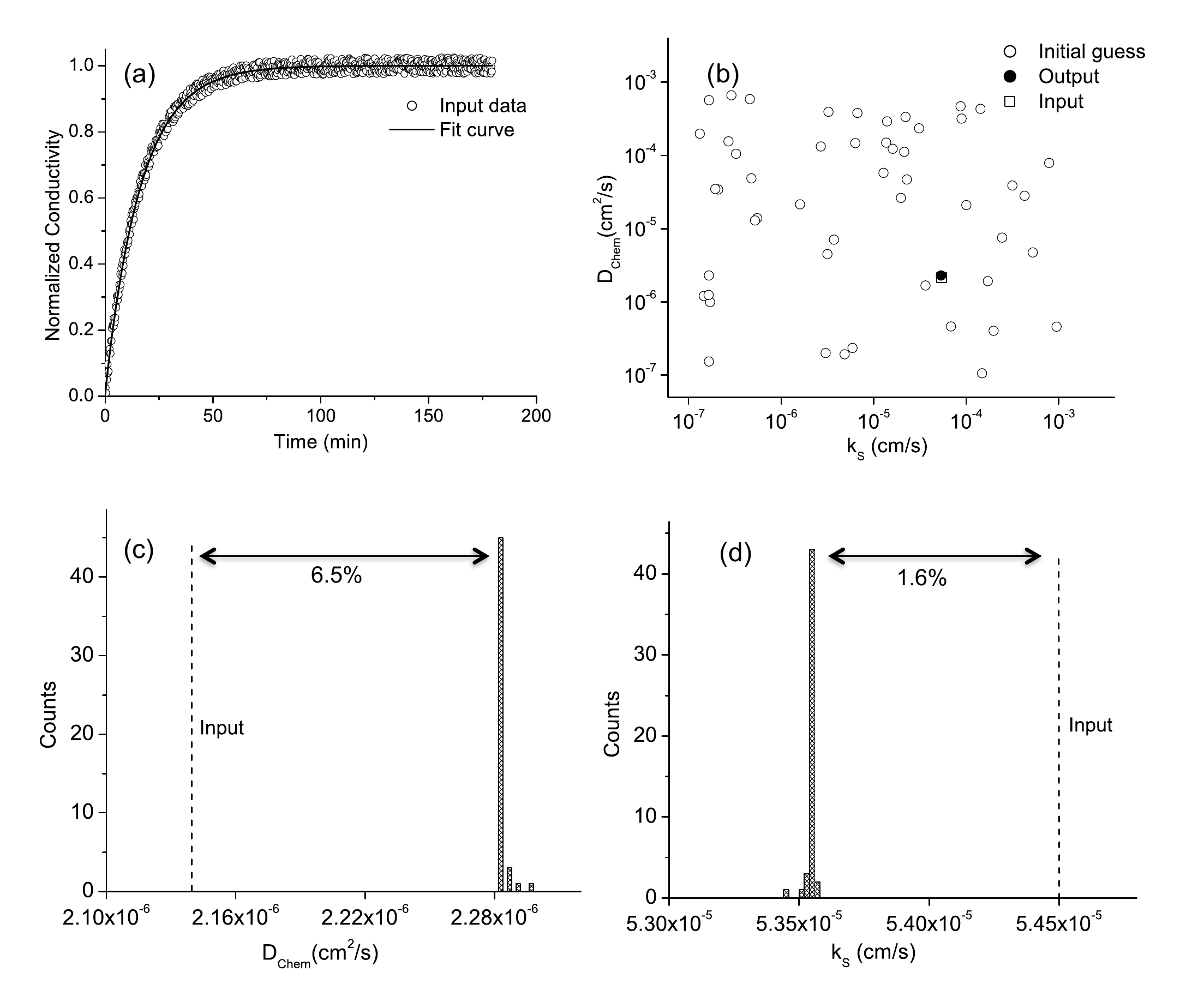}
\caption{Illustration of procedures employed to extract $D_{Chem}$ and $k_S$ from ECR data. (a) Fit to  relaxation data generated using $D_{Chem} =2.14\times10^{-6}$ cm$^2$/s, $k_S = 5.45\times10^{-5}$ cm/s and sample thickness = 0.1 cm. (b) A map of  {$D_{Chem}$, $k_S$} used as initial guess values (open circles) and the output optimized  set of values (closed circles). (c) Histogram of $D_{Chem}$ and (d) $k_S$ showing the respective mode values, $2.28\times10^{-6}$ cm$^2$/s and $5.36\times10^{-5}$ cm/s, agree well with the input values used to generate the dataset.}
\label{fig:test-fitting-code}
\end{figure*}
\begin{table*}[!t]
\small
  \caption{Representative results from testing the data analysis routine on datasets generated with  known values of sample
   thickness (2$a$ = 0.1 cm), chemical diffusion coefficient ($D_{Chem}$) and surface reaction rate constant ($k_S$). Superscript `only' indicates fits performed  using the relevant one-parameter model. }
  \label{tab:test-fitting-code}
  \centering
  \begin{tabular}{|l l l l l l l l|}
    \hline
    \multicolumn{3}{|c|}{Input}&\multicolumn{5}{|c|}{Output} \\\hline
    $\tilde{L}$ & $k_S (cm/s)$ & $D_{Chem}$ (cm$^2$/s) & $\tilde{L}$ & $k_S$ (cm/s)& $D_{Chem}$ (cm$^2$/s)& $k^{\mathrm{only}}_S (cm/s)$ & $D^{\mathrm{only}}_{Chem}$ (cm$^2$/s) \\\hline
	0.01   & $1.15\times10^{-5}$ & $4.14\times10^{-5}$ &  0.34  & $1.27\times10^{-5}$ & $1.49\times10^{-6}$ & $1.16\times10^{-5}$ & $1.60\times10^{-7}$ \\
	0.11   & $1.15\times10^{-5}$ & $4.14\times10^{-6}$ &  0.35  & $1.24\times10^{-5}$ & $1.38\times10^{-6}$ & $1.10\times10^{-5}$ & $1.57\times10^{-7}$ \\
	1.02  & $5.45\times10^{-5}$ & $2.14\times10^{-6}$ &  0.94 & $5.36\times10^{-5}$ &  $2.28\times10^{-6}$ &$4.10\times10^{-5}$ & $5.70\times10^{-7}$ \\
	10.14  & $1.05\times10^{-4}$ & $4.14\times10^{-7}$ &  10.09  & $1.04\times10^{-4}$ & $4.14\times10^{-7}$  & $2.42\times10^{-5}$ & $3.30\times10^{-7}$ \\
	99.00 & $5.05\times 10^{-4}$ & $2.04\times10^{-7}$ & 89.78& $4.59\times10^{-4}$ & $2.04\times10^{-7}$ & $1.52\times10^{-5}$ & $2.00\times10^{-7}$\\ 
    \hline
  \end{tabular}
\end{table*}
Polycrystalline compacts of SDC15 were prepared from commercial powders of the target composition  Ce$_{0.85}$Sm$_{0.15}$O$_{1.925}$ (Fuel Cell Materials Inc., Lot \#247-085, surface area=8 m$^2$/g). The powder was subjected to uni-axial pressing at 160 MPa, cold isostatic pressing at 300 MPa, followed by sintering at 1500 $^{\circ}$C for 8 h under stagnant air. Resulting samples had densities $>95\%$ of theoretical values and mean grain sizes of $\sim$3 microns, Figure~\ref{fig:sample_microstructure}. Dimensions were typically 25 $\times$ 5.5$ \times$ (0.2-2)mm$^3$. In order to ensure reproducibility of the surface characteristics, samples were polished to a final roughness of 3 $\mathrm{\mu}$m. The composition of the polished samples was confirmed by electron probe microanalysis (EPMA)  (JEOL JXA-8200, carbon coated samples, CePO$_4$ and SmPO$_4$ used as reference standards). Measurements at three different positions on a representative sample yielded absolute CeO$_2$ and Sm$_2$O$_3$ molar contents of  83.6\% $\pm$ 0.7\% and 15.3\% $\pm$ 0.9\%  respectively.\\
                                                                                                                                                                                  
To eliminate electrode contributions to the measured resistance, the conductivity was measured in a four-probe configuration. Gold electrodes were employed. Integrity of the contacts was assured by sputtering a 100 nm layer of gold at the four contact regions (208HR, Cressington, UK) and then applying an additional layer of gold by brush painting (Fuel Cell Materials, Lot \#5C149). The sample was then annealed under stagnant air at 900 $^{\circ}$C for an hour, ultimately creating porous and interconnected electrodes, as verified by SEM imaging. Gold wires were then securely wrapped around these contact points. The magnitude of the surface reaction rate constant was enhanced in some instances (to improve the accuracy of the measurement of the diffusion coefficient) by application of a layer of Pt nanoparticles to the sample surface. This was achieved by sputtering a 10 nm layer of Pt, which was then annealed for two hours at 900 $^\circ$C under stagnant air. This procedure yielded a monolayer of uniformly distributed, isolated Pt particles with average size of approximately 100 nm and average inter-particle distance of 400 nm, Figure~\ref{fig:Pt-particles}.\\

Measurements were made in an in-house constructed ECR reactor with a sample chamber approximately 1.27 cc in volume. The small size ensured rapid changes in gas-phase \textit{p}O$_2$, whereas the use of computer controlled solenoid valves ensured plug flow behavior. For measurements under relatively oxidizing conditions ($10^{-5}$ to 1 atm in $p\text{O}_2$) dry O$_2$ and Ar mixtures were used. To attain target \textit{p}O$_2$ values in the reducing regime ($p\text{O}_2<10^{-14}$ atm), mixtures of H$_2$/H$_2$O/Ar or CO/CO$_2$/Ar were employed. In the former case, the pH$_2$O was set, in all cases, at 0.023 atm by passing pre-mixed Ar and H$_2$ gases through a water bubbler held at 23 $^\circ$C. Equilibrium values of conductivity were first measured using a yttria-stabilized zirconia based oxygen sensor with an integrated s-type thermocouple for monitoring the \textit{p}O$_2$ and temperature inside the reactor. For subsequent ECR measurements, only the temperature was directly monitored and the sample conductivity was used to indicate the oxygen partial pressure, a procedure that circumvented calibration difficulties encountered during prolonged use of the sensor.\\

At each \textit{T} and \textit{p}O$_2$, ECR measurements were repeated 2-4 times. Step changes were applied in both the oxidation and reduction directions (and equivalence between the two directions confirmed). The average between the initial and final \textit{p}O$_2$ values is reported as the measurement \textit{p}O$_2$. A Keithley 2420 sourcemeter was used to measure I-V characteristics every second, from which the DC resistance was obtained. The supplied current was adjusted to vary between 1 $\mathrm{\mu A}$ and 50 $\mathrm{\mu A}$, ensuring that the potential drop across the length of the specimen was under 100 $\mathrm{mV}$. Measurements were made at 750 $^\circ$C, 800 $^\circ$C and 850 $^\circ$C. From an extrapolation of previously reported\cite{Chueh2011} grain boundary and bulk properties of SDC15 from the same supplier, the present samples with $\sim$3 $\mathrm{\mu}$m grains are expected to have a maximum grain boundary contribution to the total resistance of no more than 3\%. Thus, the relaxation behavior is justifiably taken to reflect the bulk response. Moreover, for the temperature and oxygen partial pressure regimes examined here, the concentration of defects generated in accordance with Equations~\ref{eq:defect_reaction} and \ref{eq:e-h-pair} are indeed generally small in concentration relative to the dopant concentration\cite{Lai2005}. Specifically, under the most reducing conditions examined $n = 0.3[Sm_{Ce}^\prime]$. At conditions of enhanced electron concentration, the expressions for computing the defect concentrations (and hence $D_{Chem}$) from the thermodynamic reduction data change, but analysis of the relaxation data is unmodified.

\subsection{Analysis of relaxation data}\label{subsec:num_details}
\begin{figure*}[!h]
\hspace{-0.25cm}\includegraphics[scale=.75]{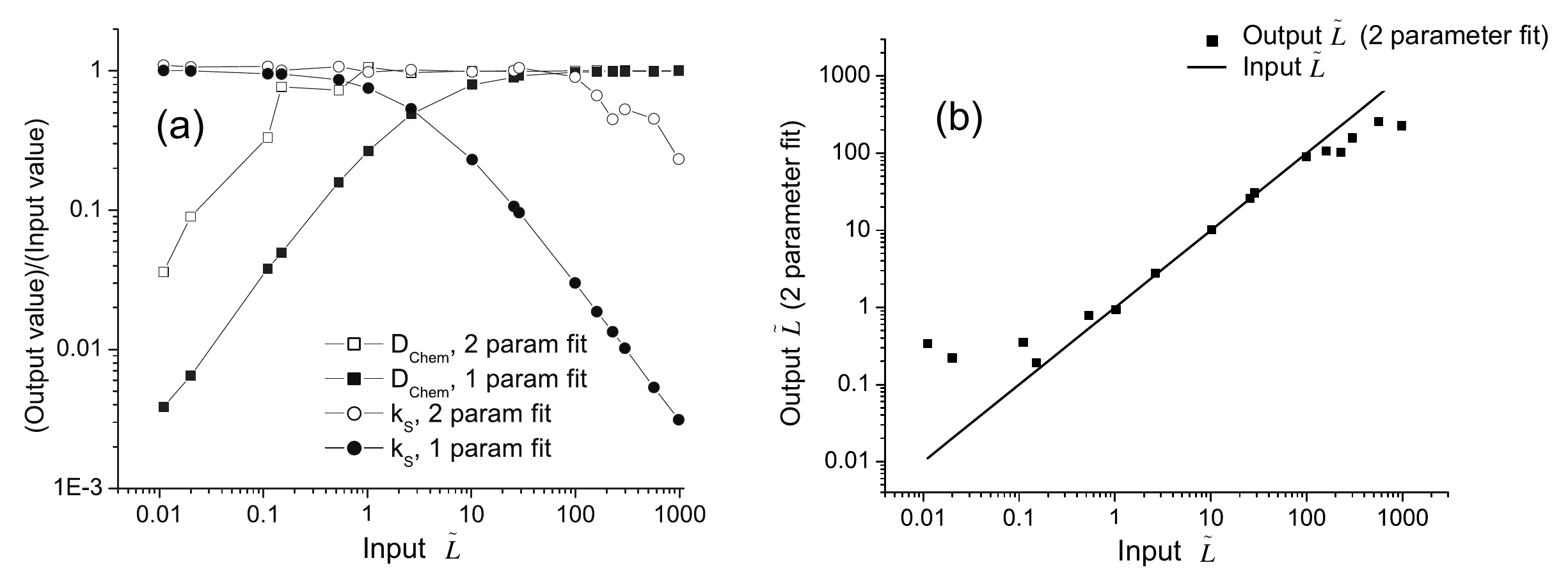}
\caption{Evaluation of numerical procedures developed for analyzing ECR data. (a) Ratio of output to input values of $D_{Chem}$ and $k_S$ as a function of input $\tilde{L}$ for the two parameter and one parameter fits, and (b) output $\tilde{L}$ from the two parameter fit plotted against input values.}
\label{fig:parameter_errors_collated}
\end{figure*}
\begin{figure}[!ht]
\hspace*{-.45cm}\includegraphics[scale=0.37]{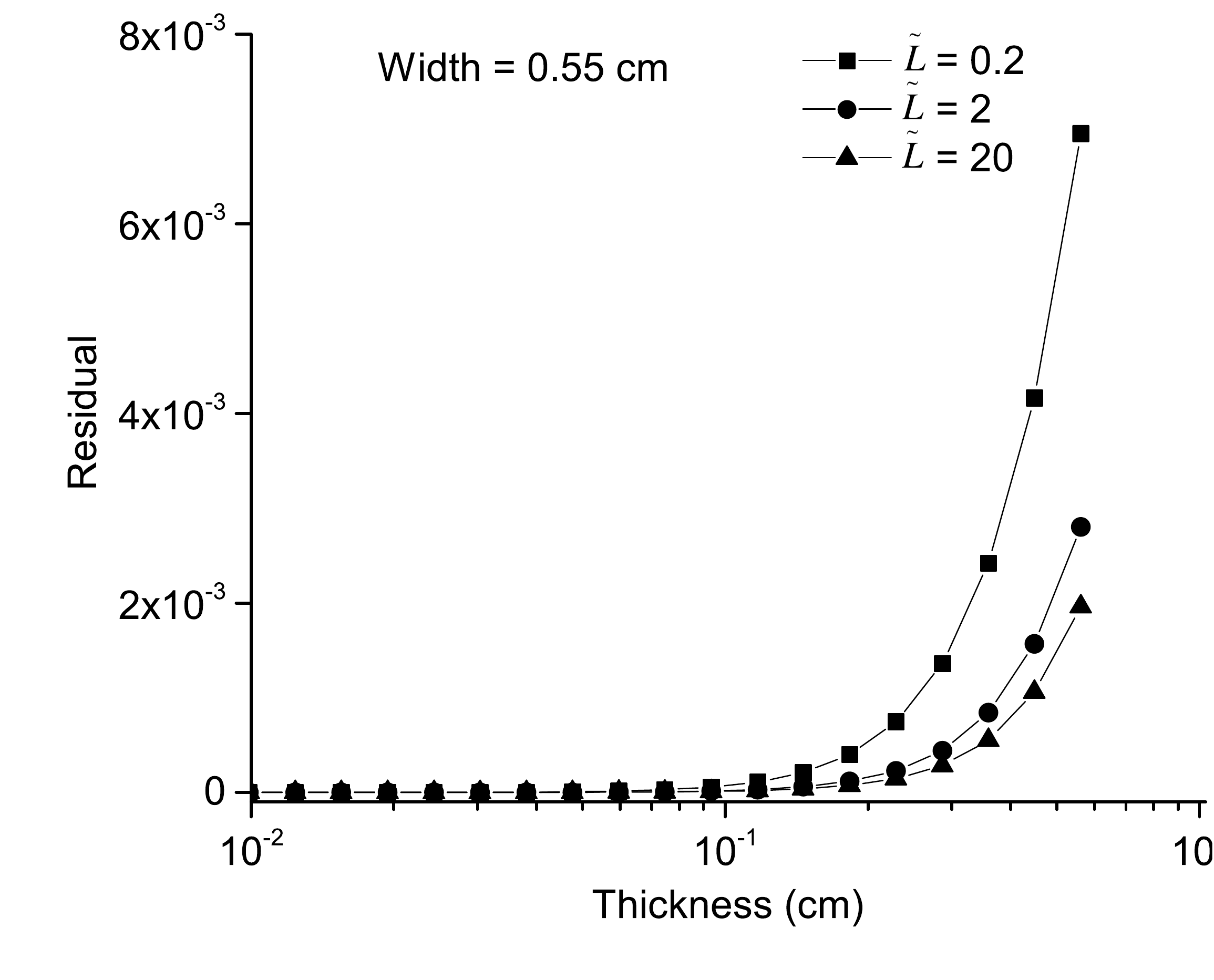}
\caption{The normalized sum of squared deviation of the 1D relaxation model from the 2D relaxation model as a function of sample thickness plotted for $\tilde{L}$ values of 0.2, 2 and 20, keeping the sample width fixed at 0.55 cm. Beyond a sample thickness of 0.15 cm, the assumption of 1D relaxation is no longer valid. }
\label{fig:residuals}
\end{figure}
The general form of the relaxation profile, Equation~\ref{eq:diffusion-surface}, can be expressed in terms of the $\alpha_m$ and $D_{Chem}$ using Equation~\ref{eq:constraint}, 
\small
\begin{multline}\label{eq:to-fit}
\dfrac{\sigma_t-\sigma_0}{\sigma_{\infty} -\sigma_0} = 1 - \sum_{n=1}^{\infty}\dfrac{2\:\tan^2(\alpha_m)}{(\alpha_m^2+\alpha_m^2\: \tan^2(\alpha_m) + \alpha_m\: \tan(\alpha_m))}\\ \exp\left({-\frac{D_{chem}\alpha_m^2t}{a^2}}\right). 
\end{multline}
\normalsize
With this formulation it is evident that there are just 2 independent parameters: $D_{Chem}$ and $\alpha_1$. The remaining $\alpha_m$ are constrained according to Equation~\ref{eq:constraint}. Guess values for $D_{Chem}$ and $k_S$ were used to obtain an initial estimate for $\tilde{L}$ and, consequently, via Equation~\ref{eq:constraint}, the set of {$\alpha_m$}. A Matlab routine was developed for then performing a constrained nonlinear fit (Equation~\ref{eq:to-fit}) to the experimental data and obtaining optimized values for $D_{Chem}$ and $k_S$. To avoid the possibility of converging to an incorrect local minimum, the procedure was repeated numerous times using randomized initial values for $D_{Chem}$ and $k_S$, each varied over 5 orders of magnitude. In the absence of significant spread, the mode of the distribution of converged estimates is reported as the experimentally derived value. It is to be emphasized that unique values for $D_{Chem}$ and $k_S$ do not necessarily imply accuracy, especially when $\tilde{L}\ll1$ or $\tilde{L}\gg1$. In these limiting cases, the same dataset was also analyzed within the framework of the simpler solutions for either surface or bulk diffusion limited processes. In all cases, it was found that the solutions converged with 3 to 4 terms included in the summation. \\

Prior to analysis of experimental data, the methodology was validated by fitting to numerically synthesized relaxation profiles, generated using given values of $D_{Chem}$ and $k_S$. For simplicity, but without any lack of generality, the sample thickness, $2a$ was fixed at 0.1 cm. Random noise with amplitude as high as 15\% was added to the generated data to simulate experimental noise. This procedure was carried out for 16 datasets, spanning $\tilde{L}$ values from $\approx$ $10^{-2}$ to $10^3$. A comparison between input and output $D_{Chem}$ and $k_S$ values provides an estimate of the errors and guidance on the preferred analysis approach, a two parameter or a single parameter fit. An example fit to simulated data is presented in Figure~\ref{fig:test-fitting-code}a, generated using input $D_{Chem}$ and $k_S$ values of $2.14\times10^{-6}$ cm$^2$/s and $5.45\times10^{-5}$ cm/s respectively, implying $\tilde{L} = 1.02$. When both diffusion and surface reaction control the relaxation rate, as in this case, the code accurately extracts both $D_{Chem}$ and $k_S$ from the data. The output of fitting using 60 different pairs of initial values for the material parameters converges towards final values that match the original input ones, Figure~\ref{fig:test-fitting-code}b. The histograms of output values of $D_{Chem}$ and $k_S$, Figures~\ref{fig:test-fitting-code}c and~\ref{fig:test-fitting-code}d, show clear peaks and minimal scatter. Furthermore, the visual quality of the fit is excellent. In this particular case, the differences between input and output values of $D_{Chem}$ and $k_S$ are 6.5\% and 1.6\%, respectively, implying that the material properties can be extracted with good accuracy.\\

Assessing, in a general manner, the confidence level that can be assigned to fit parameters is an important part of any analytical procedure. It can be readily surmised for a conductivity relaxation study that the difference between true (input) and fit (output) $D_{Chem}$ and $k_S$ values will depend on $\tilde{L}$. Specifically, when $\tilde{L}$ is large, the surface reaction step is very fast, implying it has negligible impact on the profile and errors on $k_S$ can be expected to be large. Conversely, when $\tilde{L}$ is small, the fast diffusion process has negligible impact on the profile, and errors on $D_{Chem}$ can be expected to be large. Selected results for a range of input $\tilde{L}$ are highlighted in Table~\ref{tab:test-fitting-code}, and the entire set of the results is represented in Figure~\ref{fig:parameter_errors_collated}. Figure~\ref{fig:parameter_errors_collated}a presents the ratio of output to input values of the two material parameters, and Figure~\ref{fig:parameter_errors_collated}b, a comparison between input and output values of $\tilde{L}$. The fitting is carried out using both the two-parameter and single-parameter models (Equations~\ref{eq:diffusion-surface}, ~\ref{eq:surface} and ~\ref{eq:diffusion}).\\

In general, the expectations of accuracy relative to the magnitude of the input $\tilde{L}$ are borne out, Figure~\ref{fig:parameter_errors_collated}a. When the input $\tilde{L}$ is $\sim$100 or greater, the output $k_S$ is several times smaller than the input value. Similarly, when the input $\tilde{L}$ is 0.15 or less, the output $D_{Chem}$ is many times smaller than the input $D_{Chem}$. In the high $\tilde{L}$ regions at which diffusion dominates the relaxation process, fits using the single parameter expression (Equation~\ref{eq:diffusion}) and those using the complete expression (Equation~\ref{eq:diffusion-surface}) give virtually indistinguishable values of $D_{Chem}$. Evidently, little error is introduced into $D_{Chem}$ despite the risk of overfitting of the data using the two-parameter expression. In contrast, in the low $\tilde{L}$ regions the difference between the $k_S$ values obtained from the two-parameter and the single-parameter fits is non-negligible. In the specific range examined of $\tilde{L}$ = 0.01 to 0.1, the two-parameter fit gives errors of 7-10\%  for $k_S$, whereas the single-parameter fit gives errors of 0-4\%. In this case, there is clear benefit, beyond computational efficiency, in selecting the simpler solution for analysis. Based on these results, one can conclude that a single parameter fit for only $D_{Chem}$ is appropriate when $\tilde{L}$ is 100 or greater, a two-parameter fit for both $D_{Chem}$ and $k_S$ is appropriate when $\tilde{L}$ lies between 100 and 0.15, and that a single parameter fit is appropriate when $\tilde{L}$ is 0.15 or smaller. In general, high accuracy in $k_S$ is obtained over a wider range of $\tilde{L}$ than is the case for $D_{Chem}$. \\

\begin{figure}[!ht]
\hspace*{-.9cm}\includegraphics[scale=0.4]{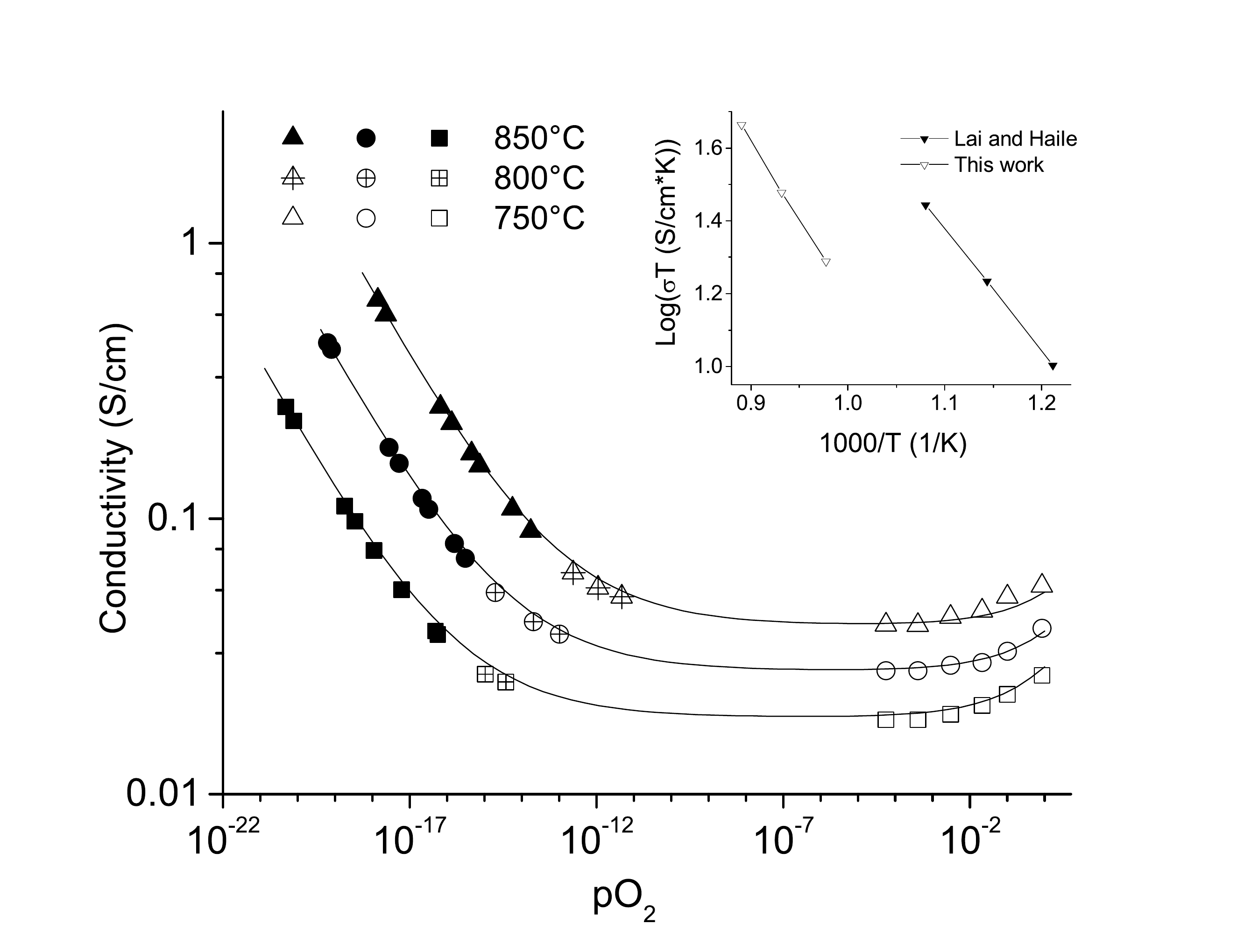}
\caption{Log-log plot of electrical conductivity of SDC15 vs \textit{p}O$_2$. Solid, cross-hair inscribed and open symbols respectively indicate data points obtained using H$_2$/H$_2$O, CO/CO$_2$ and dry O$_2$/Ar mixtures. Solid lines show fit to Equation~\ref{eq:conductivity_model_observed}. Inset is an Arrhenius plot of the ionic conductivity compared with the work of Lai and Haile\cite{Lai2005}.}
\label{fig:sdc15_conductivity}
\end{figure}

\begin{figure*}[!h]
\centering
\hspace*{-0.6cm}\includegraphics[scale=0.73]{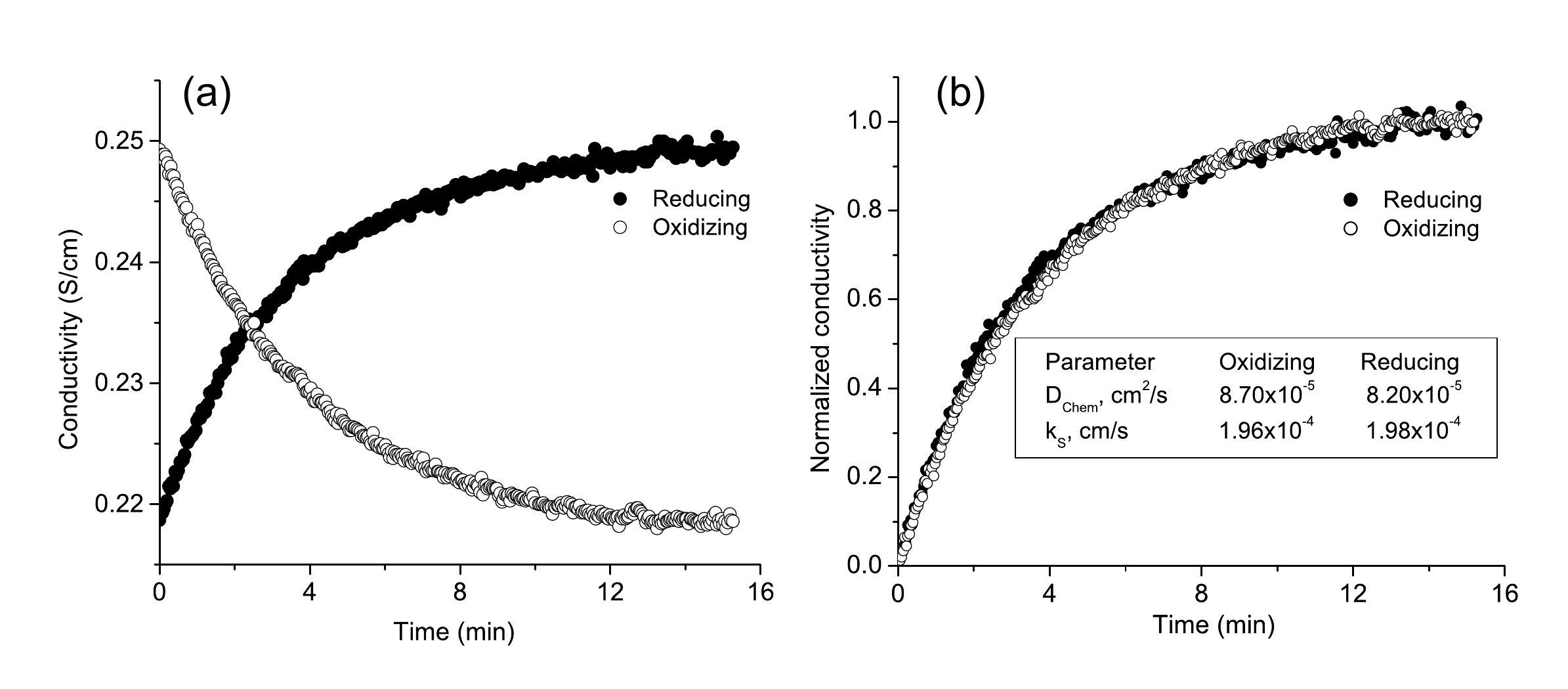}
\caption{(a) Raw conductivity relaxation profiles along reducing and oxidizing directions for a $p\text{O}_2$ switch between between $6.60\times10^{-17}$ atm and $1.33\times10^{-16}$ atm at 850 $^\circ$C. The 0.8 mm sample was sputtered with Pt catalyst particles. (b) The normalized conductivity relaxation profiles are statistically identical, confirming that the $\Delta p\text{O}_2$ is small enough to ensure the driving force, $D_{Chem}$ and $k_S$ are the same along both directions and that the system response is linear.}
\label{fig:relaxation_oxd_red}
\end{figure*}
The discussion above is framed in terms of the actual (or input) $\tilde{L}$. However, what one obtains from an analysis of experimental data is the output $\tilde{L}$. From Figure~\ref{fig:parameter_errors_collated}b, it can be seen that these two quantities are almost identical when $\tilde{L}$ lies between 0.15 and 100, consistent with the appropriateness of a two-parameter fit in this region. At the extrema, however, $\tilde{L}$ appears to plateau at $\sim$0.15 and $\sim$100.  Because the $D_{Chem}$ value obtained at high $\tilde{L}$ is insensitive to whether a two- or single-parameter fit is selected, accurate knowledge of $\tilde{L}$ is not required for accurate determination of the diffusivity. In the case of $k_S$, however, enhanced accuracy when using the single parameter fit at small $\tilde{L}$ motivates identification the appropriate formalism. From the data in Figure ~\ref{fig:parameter_errors_collated}a, it is apparent that $k_S$ from the two parameter fit is always greater than that from the single parameter fit. However, the difference between the two drops to about 5\% when the input $\tilde{L}$ is less than 0.15. This observation provides the final guidance on the how to select the fitting procedure in the absence of \textit{a priori} knowledge of the true $\tilde{L}$. Specifically, if $k_S$ (2-parameter) differs from $k_S$ (1-parameter) by less than 5\%, the latter is likely closer to the `true' value.\\

The analysis performed on this broad set of simulated data provides universal guidance on the most suitable analysis procedures for extracting $D_{Chem}$ and $k_S$ from conductivity relaxation profiles. The results in Figure~\ref{fig:parameter_errors_collated} furthermore provide an estimate of the uncertainties in the derived values when the optimal fitting procedure has been employed. It is to be emphasized, however, that if the wrong single-parameter fitting procedure is utilized, the output parameters will be almost valueless. For example, for an input $\tilde{L}$ of 0.11, a fit using only $D_{Chem}$ gives a diffusivity that is almost 30 times larger than the true value. Unless one also analyzes the data using the two-parameter methodology or can visually recognize a poor fit, the factor of 30 error could be easily overlooked. The analogous situation holds for an  evaluation of $k_S$ from a single-parameter fit at large $\tilde{L}$. Accordingly we conclude that, in the absence of \textit{a priori} knowledge of (approximate) material properties, any analysis of ECR profiles must include two-parameter fits as well as selected use of single-parameter fits in order to ensure accuracy of the output parameters.\\

As assessment of the validity of a 1D solution for the samples fabricated here was carried out by computing the difference between relaxation profiles generating using Equation~\ref{eq:diffusion-surface} and those generated using the analogous 2D expression\cite{Crank1975}. The difference is defined as
\begin{equation*}
  \sum_{n=1}^N \frac{(\tilde{\sigma}_{1\text{D}}(n\Delta t)-\tilde{\sigma}_{2\text{D}}(n\Delta t))^2}{N}
\end{equation*}
where $\tilde{\sigma}_{1\text{D}}$ and $\tilde{\sigma}_{2\text{D}}$ are the relaxation profiles generated using the 1D and 2D models respectively, $\Delta t$ is the simulation time step and $N$ is the total number of time steps. The calculation was performed for samples with thicknesses varied from $0.01$ and $0.55$ cm at selected (fixed) values of $\tilde{L}$. For generation of the profile from the 2D sample, the width was set to 0.55 cm. This brief analysis, presented in Figure~\ref{fig:residuals}, indicates that the errors in $k_S$ and $D_{Chem}$ will exceed $\sim$ 15\% when $2a$ reaches 20\% of the next largest dimension. Accordingly, samples with thicknesses greater than 0.11 cm were analyzed using the 2D solution to the diffusion equation. The 2D analysis yielded broad histograms in the output $k_S$ and $D_{Chem}$ values and in contrast to the 1D analysis, the modes of these distributions did not correspond to the solution with the minimum least squared error. For these cases, the latter are reported as the experimentally derived values.\\

\section{Results and Discussion}\label{sec:results}
 \subsection{Equilibrium conductivity}
\begin{figure}[!h]
\hspace*{-.3cm}\includegraphics[scale=.4]{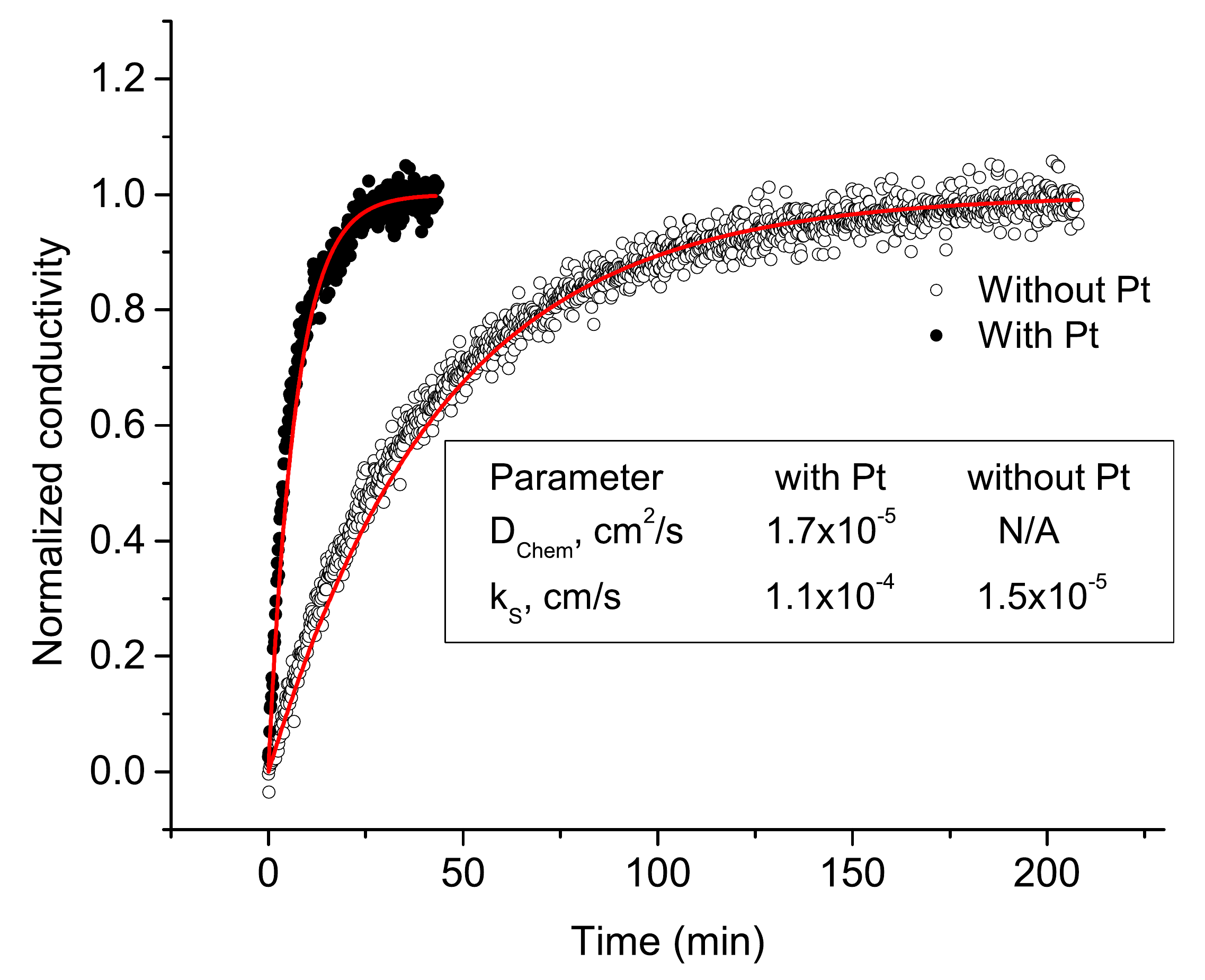}
\caption{Relaxation profile of 0.8 mm thick SDC 15 sample with and without Pt catalyst on the surface for identical measurement conditions. $T=750$ $^\circ$C, $p$H$_2$ = 0.1 atm, $p$H$_2$O = 0.023 atm, balance Ar. $\Delta p\text{O}_2=6.0\times10^{-21}$atm to $2.0\times10^{-21}$atm. The solid red lines are fit profiles. Without Pt, only the slow surface reaction step could be measured. }
\label{fig:relaxation_with_without_Pt}
\end{figure}
Figure~\ref{fig:sdc15_conductivity} shows the \textit{p}O$_2$ dependence of the total electrical conductivity of SDC15 at 750, 800 and 850 $^\circ$C, with relevant transport parameters summarized in Table~\ref{tab:conductivity_comparison}. Under reducing conditions (low \textit{p}O$_2$), the total conductivity is predominantly electronic, showing the expected n-type behavior with a -0.25 power law dependence on \textit{p}O$_2$. Morever, the value of the n-type conductivity is in excellent agreement with earlier results from Lai \cite{Lai2005} and from Chueh\cite{Chueh2008} reported from similar starting materials. With increasing \textit{p}O$_2$, the conductivity plateaus to a constant value, reflecting the occurrence of the electrolytic regime. At the highest values of \textit{p}O$_2$, the total conductivity increases, indicating the onset of p-type conductivity. However, the power law dependence is found to be best described with an exponent of 0.35 rather than the expected value of 0.25. The solid lines in the figure reflect a fit to the expression
\begin{equation}\label{eq:conductivity_model_observed}
\sigma_{tot} = \sigma_n^0p\text{O}_2^{-0.25} +\sigma_{ion} + \sigma_p^0p\text{O}_2^{0.35}
\end{equation}
rather than to Equation~\ref{eq:conductivity_model_expected}, and it is evident the data are well-represented by this expression. The ionic conductivity derived from the fit is shown in the inset of Figure~\ref{fig:sdc15_conductivity}.
\begin{table}[!h]
\small
\caption{Parameters describing the conductivity of SDC15, based on a fit of the expression in Equation~\ref{eq:conductivity_model_observed} to determine the ionic, n-type, and p-type conductivities, as well as fit to an Arrhenius expression ($\sigma T = A\exp(-E_a/k_bT)$)}
\label{tab:conductivity_comparison}
\centering
\begin{tabular}{|l|l|l|l|}
\hline & $E_a$, eV & A, S/cm K & $\sigma$ (800 $^\circ$C) S/cm\\ \hline
Ionic & 0.85 & $2.95\times10^5$& 0.029 \\ 
n-type & 2.35 & $7.6\times10^8$ & 0.222 (\textit{p}O$_2=10^{-18}$ atm) \\ 
p-type & 0.22 & $2.18\times10^2$ & 0.051 (\textit{p}O$_2=1$ atm)\\\hline
\end{tabular}
\end{table}

In contrast to the n-type conductivity, the ionic conductivity measured here is lower, by about a factor of three, than that obtained earlier by Lai for SDC15 \cite{Lai2005} (see inset). The activation energy for ionic transport obtained here is, however, consistent with typical bulk values\cite{Balazs1995}, supporting the statement that grain boundary influences on the relaxation behavior are negligible. The difference between previous and present measurements is tentatively attributed to the differences in source materials (although the powders were from the same supplier, they were of different types, nanocrystalline versus microcrystalline),  as well as slightly different pellet fabrication procedures, with a more aggressive sintering protocol having been employed here in order to obtain large grains. A comparable level of scatter in the literature has been noted by Mogensen et al. for 20 mol\% Sm and Gd doped ceria \cite{Mogensen2000}. In that case, the scatter was hypothesized to originate from differences in grain boundary contributions to the total resistance. The microstructure of the present samples renders such an explanation unlikely to be applicable in this work (as the grains are large enough to render the grain boundary contribution negligible, as discussed above). Nevertheless, the low number density of grain boundaries in the materials studied here can be conceived to influence the impurity levels in the bulk and, through that avenue, plausibly influence the bulk ionic conductivity. It is to be emphasized that the EPMA results show the Sm doping level to match the nominal value of 15 mol\%, and thus a reduced dopant level cannot be responsible for the reduced ionic conductivity.\\
\begin{figure}[!ht]
\hspace*{-0.4cm}\includegraphics[scale=0.4]{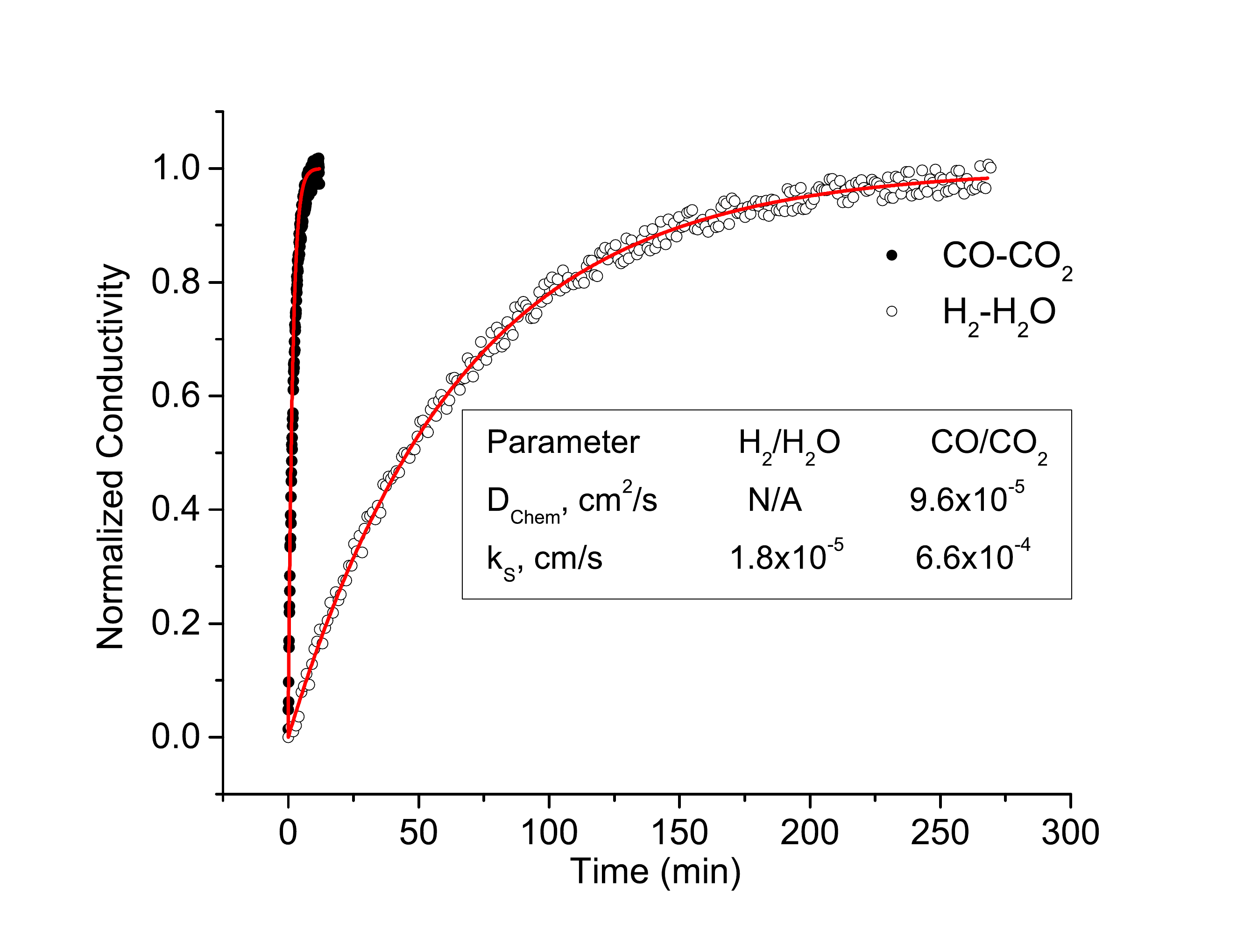}
\caption{Relaxation profiles of a 1.72 mm thick SDC 15 sample at 800 $^\circ$C, $p\text{O}_2=2.3\times10^{-15}$ atm using H$_2$/H$_2$O ($\Delta p\text{O}_2=3.0\times10^{-15}$ atm to $1.7\times10^{-15}$ atm) and $p\text{O}_2=2.2\times10^{-13}$ using CO/CO$_2$ ($\Delta p\text{O}_2=3.4\times10^{-13}$ atm to $1.0\times10^{-13}$ atm) mixtures. The solid red lines are fit profiles. Although $D_{Chem}$ is slightly higher under the more oxidizing conditions of the CO/CO$_2$ experiment, the dramatically enhanced relaxation rate is largely a result of the differences in $k_S$. }
\label{fig:relaxation_COCO2_H2H2O}
\end{figure}   
\begin{figure}[!h]
 \centering
\hspace*{-0.3cm}\includegraphics[scale=0.33]{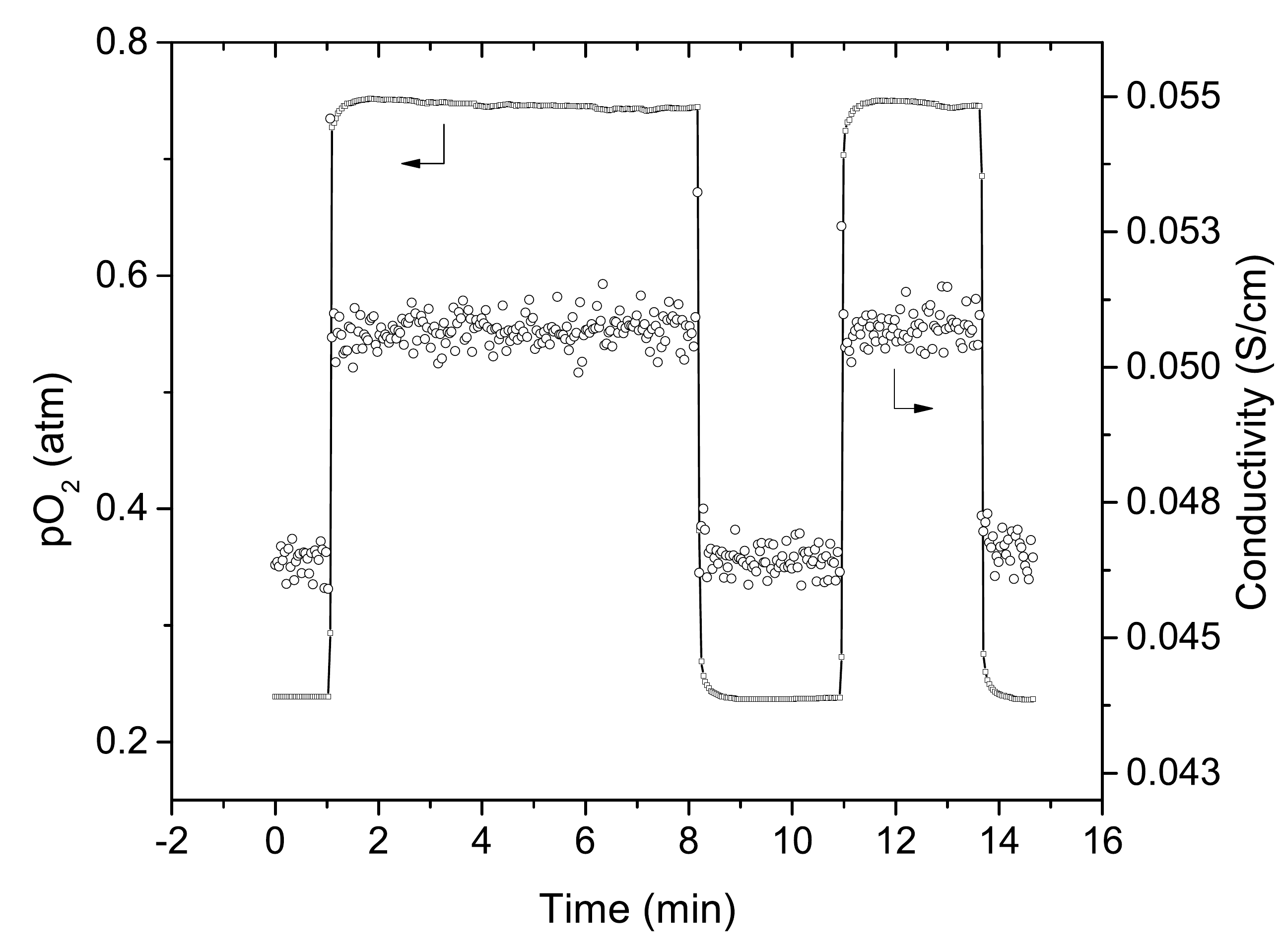}
\caption{Electrical conductivity and \textit{p}O$_2$ as a function of time for a step change $\Delta p\text{O}_2$: $2.6\times10^{-1}$ atm and $7.9\times10^{-1}$ atm (p type behavior) at 850 $^\circ$C. A 0.8 mm thick sample without Pt catalyst on the surface shows dramatically fast re-equilibration times, less than 5 seconds. Also, note the \textit{p}O$_2$ switch times of 1 to 2 seconds.}
\label{fig:p-n-type}    
\end{figure}  
\subsection{Relaxation Behavior}\label{sec:ECRresults}
Example relaxation profiles are presented in Figure~\ref{fig:relaxation_oxd_red} for a measurement carried out under reducing conditions in a H$_2$-H$_2$O-Ar mixture at 850 $^\circ$C using a Pt-catalyzed sample 0.8 mm in thickness ($a$ = 0.4 mm), in both the oxidizing and reducing directions. It is apparent that the forward and reverse directions yield normalized conductivity profiles that are statistically identical, confirming that the step change between $6.6\times10^{-17}$  and $1.3\times10^{-16}$  atm was small enough to justify the assumptions of the analytical procedure.\\

The dramatic influence of Pt nanoparticles alluded to above on the relaxation process is evident in Figure~\ref{fig:relaxation_with_without_Pt}. In the absence of Pt, the relaxation time for the step change reflected in Figure~\ref{fig:relaxation_oxd_red} increased from $\sim$20 to $\sim$200 min, and  $\tilde{L}$ decreased from 0.28 to a value less than 0.1, motivating an analysis according to Eq.~\ref{eq:surface} for a process entirely limited by the surface reaction step. As described above, it was anticipated, based on the reported values of $D_{Chem}$ and $k_S$ under H$_2$/H$_2$O/Ar mixtures, that SDC samples of the dimensions utilized here would be surface reaction limited. That Pt, which can only influence $k_S$, enhances the relaxation rate directly confirms the expectation of a surface reaction limited process. A consequence of the relatively slow surface reaction kinetics on bare SDC15 is the inaccessibility of $D_{Chem}$ from samples thin enough to retain the validity of the 1-dimensional approximation. Rather than increase $a$ to achieve $\tilde{L} \geq 0.15$, an adjustment which would have dramatically increased the measurement time, all measurements of $D_{Chem}$ under H$_2$-H$_2$O-Ar mixtures were carried out using Pt catalyzed samples. While elucidation of the mechanisms by which Pt catalyzes the dissociation/formation of H$_2$O on the surface of doped ceria is beyond the scope of this study, we note that Wang et al. have recently reported a similar enhancement in conductivity relaxation rates in doped ceria in the presence of Pt nanoparticles\cite{Wang2013}. More generally, it is widely recognized that precious metal particles on ceria supports form a highly active combination for catalyzing a broad range of chemical reactions\cite{Trovarelli1996}. The ECR method provides a rigorous approach for studying these phenomena.\\

Additional evidence for the major role of surface reaction kinetics in the relaxation behavior of SDC15 samples of moderate thickness (specifically 1.72 mm) under reducing conditions is presented in Figure~\ref{fig:relaxation_COCO2_H2H2O}, in which the profiles of the bare oxide under H$_2$-H$_2$O-Ar and CO-CO$_2$-Ar at 800 $^\circ$C are compared. Although $D_{Chem}$ is slightly larger under the more oxidizing conditions of the CO-CO$_2$-Ar experiment, \textit{p}O$_2=2.2\times10^{-13}$ vs. \textit{p}O$_2$ = $2.3\times10^{-15}$ atm, the observed 10-fold reduction in relaxation time is, by far, a result of the increased surface reaction rate. A fit to the relaxation data reveals that the $k_S$ in the CO-CO$_2$-Ar mixture is a remarkable $\sim$40 times greater than it is in the H$_2$-H$_2$O-Ar mixture (an order of magnitude greater than it is on Pt-catalyzed SDC15 in H$_2$-H$_2$O-Ar). Again, studying the catalytic behavior of SDC is beyond the scope of this paper, but these preliminary data immediately suggest that thermochemical production of CO will be kinetically favorable over H$_2$ production. Furthermore, from the perspective of ECR experimental design, the rapid surface exchange enables ready measurement of $D_{Chem}$ in the intermediate \textit{p}O$_2$ region accessible using CO-CO$_2$-Ar mixtures without the need for a catalyst. Conversely, whereas an initial evaluation of literature values of $k_S$ and $D_{Chem}$ indicated these experiments would be well within the surface reaction limited regime, a co-limited process is clearly encountered under CO/CO2 mixtures. This result highlights the importance of evaluate the data in an unbiased manner, without presupposing the nature of the experimental regime.\\

An example relaxation profile under oxidizing conditions is presented in Figure~\ref{fig:p-n-type} for a sample 0.8 mm in thickness as measured at 850 $^\circ$C. At the outset it was anticipated, as discussed above, that measurements under these conditions would be difficult due to the low sensitivity of total conductivity to \textit{p}O$_2$ in this regime. However, changes in conductivity between start and finish of the relaxation of $\sim$3\% are evident and readily recorded, consistent with the enhanced electronic transference numbers of the SDC15 employed here. At a \textit{p}O$_2$ of $0.5$ atm, the electronic contribution to the transport is p-type, as evidenced by the increase in conductivity with increasing \textit{p}O$_2$ and also directly indicated by the equilibrium conductivity results, Figure ~\ref{fig:sdc15_conductivity}. A striking feature of these profiles is the exceptionally fast response time of 10 s, approaching the reactor flush time of 1 to 2 s and precluding the extraction of meaningful kinetic parameters. The sample thickness had to be increased to 1.72 mm to sufficiently slow the relaxation kinetics and enable acquisition of useful data (not shown). Although these thicker samples required analysis according to the 2-dimensional solution (Figure~\ref{fig:residuals}) and accordingly substantially longer computing times, both $k_S$ and $D_{Chem}$ could be reasonably determined.\\

\begin{figure}[!h]
\hspace{-0.4cm}\includegraphics[scale=0.39]{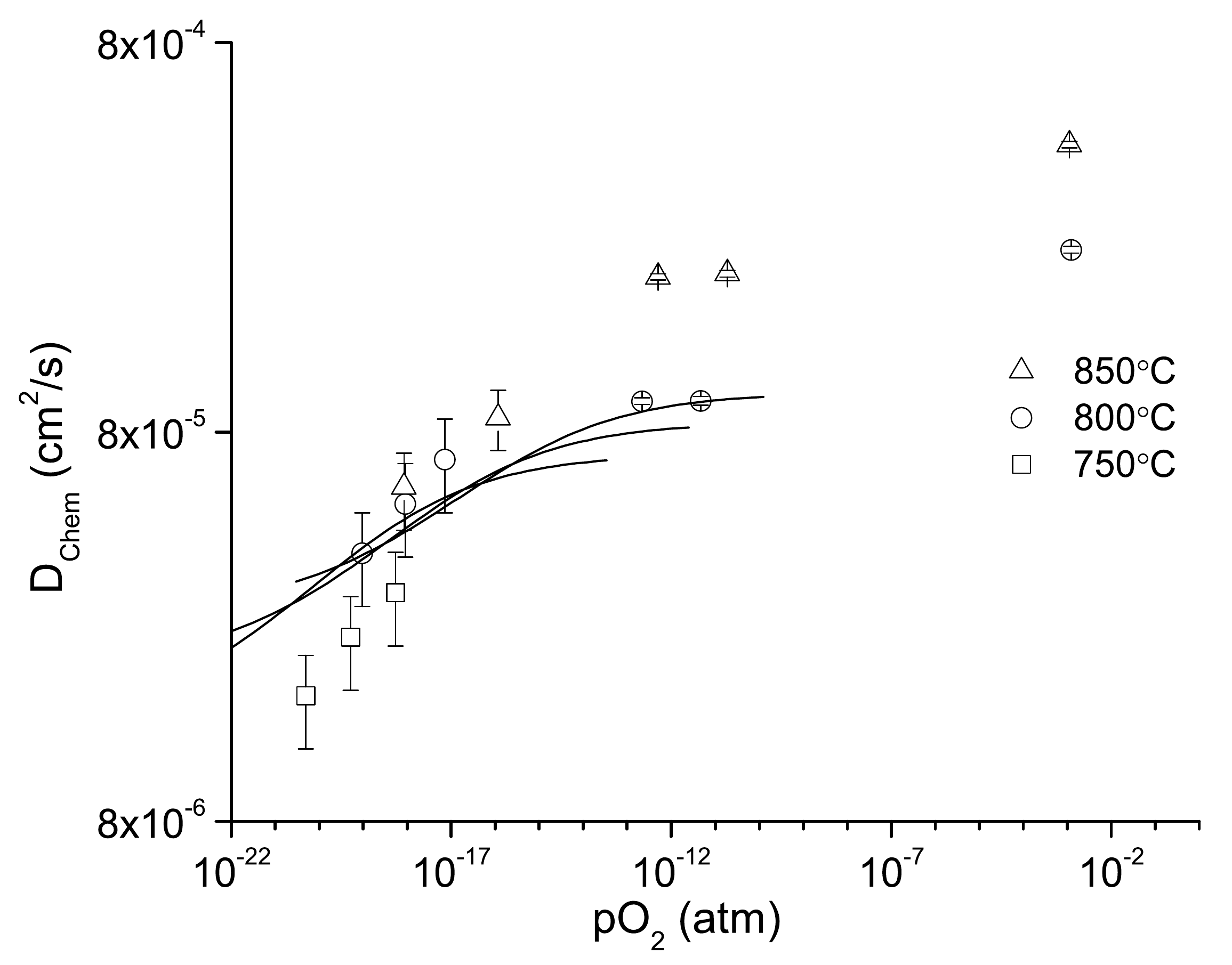}
\caption{$D_{Chem}$ as a function of \textit{p}O$_2$ at 750 $^\circ$C, 800 $^\circ$C and 850 $^\circ$C from this study overlaid on approximate analytical values computed  assuming an ideal solution model (computed values based on  extrapolations of defect concentrations and mobilities measured at lower temperatures\cite{Lai2005}).}
\label{fig:D-expt}
\end{figure}

 \begin{figure}[!h]
\hspace{-0.3cm}\includegraphics[scale=.39]{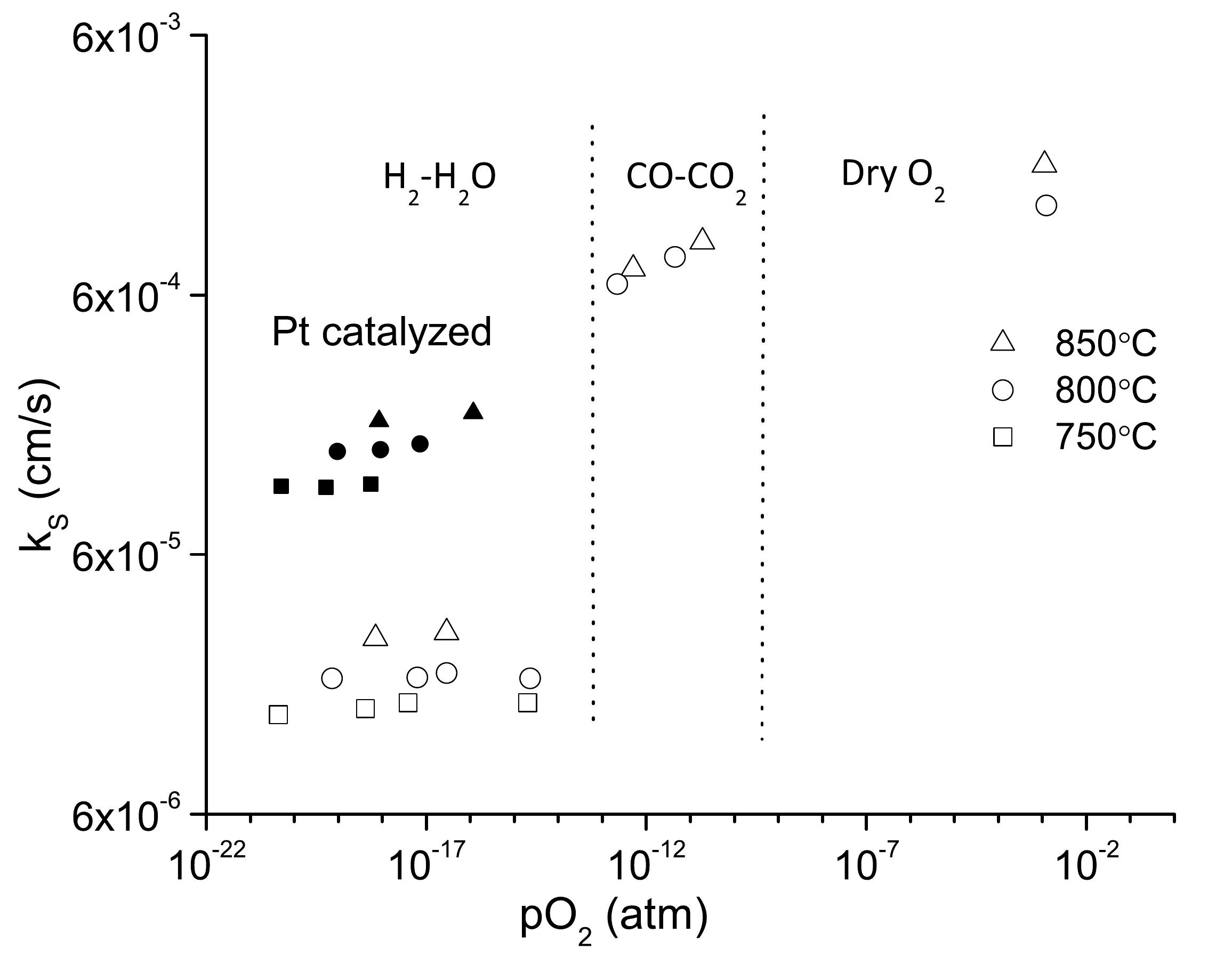}
\caption{$k_S$ as a function of \textit{p}O$_2$ at 750 $^\circ$C, 800 $^\circ$C and 850 $^\circ$C from this study. The abrupt jump in $k_S$ at intermediate \textit{p}O$_2$ corresponds to a change in the gas mix from H$_2$/H$_2$O to  CO/CO$_2$. While $D_{Chem}$ is dependent only on \textit{p}O$_2$, $k_S$ shows a much stronger dependence on the gas species.}
\label{fig:k-expt}
\end{figure}   

The diffusivity results obtained from these experiments are summarized in Figure~\ref{fig:D-expt} with errors, which represent the minimum fitting errors, estimated from the analysis presented in Figure 4. The directly measured values are compared to those computed using the conductivities presented in Figure 6 and thermodynamic properties reported by Lai and Haile\cite{Lai2005}. Overall, the agreement is satisfactory, validating the methodology. Under the most reducing conditions of this study, at which the defect concentrations are dominated by the Brouwer approximation of Equation~\ref{eq:electroneutrality}, but conductivity is n-type, $\sigma_{ion} <\sigma_n$ and $c_{ion} > c_n$, leading to a $D_{Chem}$ that is inversely proportional to $c_n$ and hence decreases with decreasing \textit{p}O$_2$. Under moderately oxidizing conditions (\textit{i.e.}, the electrolytic regime), although $D_{Chem}$ becomes difficult to measure by ECR, its behavior can be described. In this region, $\sigma_{ion} > \sigma_n$ and $c_{ion} \gg c_n$, and thus $D_{Chem}$ asymptotes to $D_{n} = \frac{RT}{F^2}\frac{\sigma_n}{c_n}$. Under these conditions the minority carrier dominates the ambipolar diffusion process. The slight deviation between experiment and calculation under moderate to low \textit{p}O$_2$ may be the result of a small dependence of the electronic mobility on oxygen partial pressure, as suggested elsewhere\cite{Chueh2010}. The very weak dependence of $D_{Chem}$ on temperature is a direct result of the competing temperature dependences of mobility and defect concentrations. A significant feature of Figure 10 is the very large $D_{Chem}$ measured when the electronic conductivity is p-type, about a factor of 3 larger than when it is n-type. When holes become the dominant minority carrier, $D_{Chem}$ can be expected to approach $D_{p}$ rather than $D_{n}$, implying that the higher chemical diffusivity is a result of the higher mobility of holes over electrons. The hole mobility can be roughly estimated using the expression ,
\begin{equation}
\mu_p = \frac{eD_p}{k_BT}
\end{equation}
 which yields a value of $\sim 2\times10^{-3}$ cm$^2$V$^{-1}$s$^{-1}$ at 800 $^\circ$C, approximately 50\% greater that the electron mobility (obtained from an extrapolation of the data published by Lai and Haile\cite{Lai2005}). Reliable values of hole mobility in rare-earth doped ceria are unavailable from the literature due to the difficulty of accurately of determining the hole concentration (the latter is required for determining mobility from a measurement of hole conductivity). We suggest that the hole mobility exceeds that of the electrons because of the delocalized nature of the O 2p band (the origin of the holes). In contrast, the electrons are localized in the Ce 4f states\cite{Silva2007}, effectively behaving as polarons, and hence are less mobile. Using the estimated mobility and the measured conductivity, the hole concentration can further be estimated, and the resulting value is $4.2\times 10^{19}$ cm$^{-3}$ at 800 $^\circ$C and 1 atm oxygen partial pressure. This concentration is equal to the electron concentration that appears at the same temperature and an oxygen partial pressure of $1.2\times10^{-14}$ atm. This rough analysis indicates that the mobilities and concentrations of holes required for explaining the high diffusivities and observed p-type conductivity are reasonable. \\

The surface reaction rate data, summarized in Figure~\ref{fig:k-expt}, are striking. As already noted, the overall magnitude of $k_S$ obtained under H$_2$-H$_2$O-Ar mixtures is generally consistent with what has been observed in the literature. However, $k_S$ decreases slightly with decreasing \textit{p}O$_2$. Our results thus not only contradict the results obtained from electrochemical measurements carried out at slightly lower temperatures, but also the general expectation that surface reaction rates increase with increasing vacancy concentration. On the other hand, the data seem to obey the often noted correlation between $D_{Chem}$ and $k_S$\cite{Lane2000a}. Most significantly, the surface reaction constant is more than two orders of magnitude higher under CO-CO$_2$-Ar and O$_2$-Ar than it is when H$_2$O is present. Again, there is some precedence for such a result, with Yashiro and coworkers also having seen a higher $k_S$ for ECR measurements under CO-CO$_2$-Ar than under H$_2$-H$_2$O-Ar\cite{Yashiro2002}, however, the underlying mechanisms that lead to this behavior remain to be elucidated. It is further noteworthy that $k_S$ values under CO-CO$_2$-Ar and under O$_2$-Ar mixtures are very similar, despite dominance of electrons as the minority carriers in the former case and holes in the latter. Another surprising result is the very weak temperature dependence of $k_S$, irrespective of gas atmosphere. Overall, this rich set of behaviors sets the stage for employing ECR methods to fully explore and understand the catalytic properties of ceria and its derivatives.

\section{Conclusions}
\label{sec:conclusions}
We have evaluated the oxygen transport properties of bulk samples of SDC15 over a wide range of \textit{p}O$_2$ at 750 $^\circ$C, 800 $^\circ$C and $850^\circ$C using electrical conductivity relaxation. SDC was chosen as a benchmarking material to demonstrate the versatility and robustness of numerical procedures developed to directly extract both bulk chemical diffusivity and surface reaction rate constant. The methodology is proven to be sound, provided the sample geometry and microstructure are tailored to justify the approximations of the analytical approach.\\

Beyond method validation, several new insights are afforded by this study of SDC. The slightly enhanced p-type conductivity of the SDC15 employed here enables ECR measurements under oxidizing conditions, and we find that $D_{Chem}$ is substantially higher in the p-type region than it is in the n-type. Both results, the high p-type conductivity and the high $D_{Chem}$, point towards much higher hole than electron mobility. The surface reaction constant in SDC is highly dependent on the nature of the gaseous species. Relative to CO-CO$_2$-Ar mixtures, H$_2$-H$_2$O-Ar mixtures appear to have a poisoning effect on the surface of SDC. The rate constants on bare SDC15 in the presence of H$_2$O are more than two orders of magnitude lower than they are in its absence. The rapid surface reaction kinetics under CO-CO$_2$-Ar mixtures suggests kinetic advantages for the production of CO rather than H$_2$ in a two-step thermochemical process.  The combination of extremely high $D_{Chem}$ and extremely high $k_S$ under relatively oxidizing conditions (leading to extremely short relaxation times) suggests the possibility of using SDC as a \textit{p}O$_2$ sensor in oxygen-rich environments.\\ 

For all cases examined in this study (samples several hundred $\mu$m in thickness) the relaxation was either largely or entirely surface reaction limited. In both thermochemical and fuel cell electrode applications, SDC is employed in a morphology with short diffusion distances, several to several tens of microns, suggesting that surface reaction limitations will dominate the performance of real devices. Accordingly, efforts at understanding and enhancing surface reaction kinetics will be essential for advancing these technologies.

\section{Acknowledgments}
This material is based upon work supported by the National Science Foundation under Grant No. CBET-1038307.





\bibliographystyle{rsc} 
\bibliography{bib_paper1} 

\end{document}